\documentclass[11pt]{article}

\usepackage{arxiv}
\usepackage[utf8]{inputenc}
\usepackage[T1]{fontenc}
\usepackage{url}
\usepackage{booktabs}
\usepackage{amsmath}
\usepackage{amsfonts}
\usepackage{amssymb}
\usepackage{microtype}
\usepackage{graphicx}
\usepackage{float}
\usepackage{algorithm}
\usepackage{algpseudocode}
\usepackage[table]{xcolor}
\usepackage{array}
\usepackage{threeparttable}
\usepackage{hyperref}

\definecolor{tabfirst}{rgb}{1.00,0.78,0.78}
\definecolor{tabsecond}{rgb}{1.00,0.88,0.70}
\definecolor{tabthird}{rgb}{1.00,1.00,0.75}
\definecolor{tabneutral}{rgb}{0.91,0.94,0.98}

\newcommand{\strongcell}[1]{\cellcolor{tabfirst}#1}
\newcommand{\suggestcell}[1]{\cellcolor{tabsecond}#1}
\newcommand{\weakcell}[1]{\cellcolor{tabthird}#1}
\newcommand{\neutralcell}[1]{\cellcolor{tabneutral}#1}

\title{FOMO: FLEX Options and Market Outcomes}

\date{}

\author{
 Shivesh Prakash$^{1*}$\href{https://orcid.org/0009-0007-4120-0921}{\includegraphics[width=10pt]{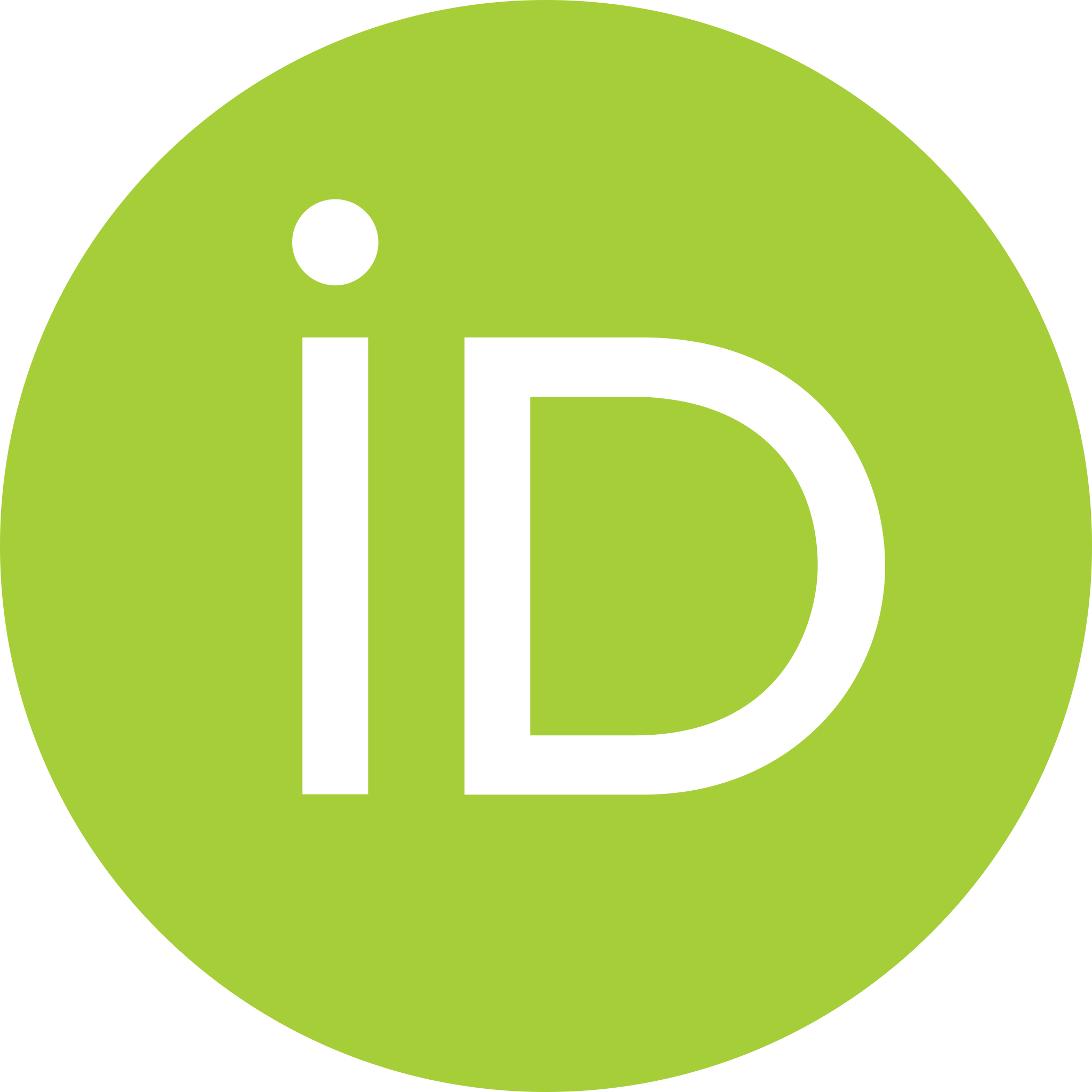}}\quad
 Frank Ferstler$^{2}$\quad
 Nicholas Pezolano$^{2}$ \\
 {\normalfont $^1$Haas School of Business, University of California, Berkeley, 2220 Piedmont Ave, Berkeley, CA 94720} \\
 {\normalfont $^2$Newmark Risk, New York, NY 10005} \\
 $^{*}$\texttt{shivesh\_prakash@berkeley.edu}
}

\begin{document}
\maketitle

\begin{abstract}
FLEX options are customized exchange-traded contracts allowing participants to choose nonstandard strikes, expirations, exercise styles, and settlement terms while retaining central clearing. We investigate whether single-name FLEX activity leaves measurable footprints in regular listed-option markets and underlying equities. Using a linked dataset of FLEX consolidations, daily open-interest files, listed-option chains, equity returns, and corporate actions over 250 trading days (March 2025--March 2026) across 41 underlyings and 53 event dates, we document three primary empirical findings: (1) Matched-control difference-in-differences estimates reveal a robust post-event increase in listed-option open interest ($\beta = 0.3061$, $p = 0.0042$, Benjamini--Hochberg $p = 0.0168$). (2) FLEX activity is strongly selected into high implied-volatility (IV) environments, concentrated in mid-price underlyings (date-fixed-effect ATM IV premium $\beta = 0.1235$, $p = 0.0030$). (3) Event-day IV-minus-realized-volatility spreads predict execution-stressed short-straddle returns ($\beta = 0.0366$, $p = 0.0100$). Short-lived equity return effects ($\text{CAR}_0$ difference of $-0.85$\%, $p = 0.0442$) do not persist over 1-to-5-day horizons. Overall, FLEX activity serves primarily as an observable marker of institutional volatility demand and open-interest positioning rather than a directional stock-return predictor.

\vspace{0.5em}
\end{abstract}

\section{Introduction}

Exchange-listed options are standardized by design. They trade at exchange-defined strikes and expirations, clear through the Options Clearing Corporation, and generally support continuous public quoting. FLEX options occupy a different institutional niche. They are exchange-traded options whose contract terms can be customized by the counterparties, including expiration date, exercise price, exercise style, and settlement features \cite{cboe_flex}. In effect, FLEX options borrow customization from over-the-counter derivatives while preserving exchange trading, clearing, and rule-based reporting.

The economic importance of this market has increased. Cboe reports that FLEX option volume grew from roughly 60,000 contracts per day in 2014 to nearly 800,000 contracts per day in 2024, a 26\% annual growth rate, compared with 7.7\% annual growth for total listed option volume over the same period \cite{cboe_flex_appeal}. FLEX options are used in institutional hedging, structured products, defined-outcome ETF construction, stock-loan and conversion/reversal applications, and payoff designs that require dates or strikes not available in the standardized option chain \cite{cboe_flex, cboe_flex_appeal, garcia_feijoo_silverstein_2023}. Despite this growth, there is limited empirical evidence on whether FLEX activity affects, or at least predicts, conditions in the regular listed-option and equity markets of the same underlying.

The absence of a large empirical literature is itself informative. Standard option-market research is usually built on continuously quoted option chains, standardized option symbols, or exchange-level order imbalance. FLEX options are different along every dimension that makes empirical work convenient: contract terms are customized, activity can be episodic, identifiers must be mapped back to the underlying, and the economic exposure may be created for institutional or product-design reasons rather than for directional speculation. A researcher must therefore combine files that are often studied separately: FLEX consolidations, daily FLEX positions, standardized listed-option aggregates, equity prices, and corporate-action adjustments. To our knowledge, public academic work has studied adjacent topics such as demand-based option pricing, option price discovery, dealer gamma, and defined-outcome funds, but has not yet provided a direct empirical event-study of single-name FLEX activity and its spillovers to regular option and equity markets. This paper is intended as an initial map of that missing object.

This gap is surprising because the regular options literature provides several reasons to expect spillovers. First, net option demand can move implied-volatility surfaces when intermediaries face inventory and hedging constraints \cite{bollen_whaley_2004, garleanu_2009}. Second, options can reveal informed trading or volatility information before it is visible in the stock market \cite{easley_1998, pan_poteshman_2006, ni_pan_poteshman_2008, johnson_so_2012, an_ang_bali_cakici_2014}. Third, option-implied and realized volatility spreads have been linked to option returns and expected stock returns \cite{coval_shumway_2001, goyal_saretto_2009, bali_hovakimian_2009}. Finally, option gamma and dealer hedge rebalancing can influence stock-return dynamics and volatility, especially when liquidity is limited \cite{avellaneda_lipkin_2003, barbon_buraschi_2021, soebhag_2023}.

FLEX options provide a useful setting because the event is not just ordinary option volume. A FLEX consolidation identifies a customized institutional contract that becomes reportable with explicit terms: underlying, call/put type, strike, expiration, and effective date. If dealers hedge such contracts in the listed-option market or the stock, FLEX effective dates should leave measurable traces. If FLEX users select names whose regular listed markets are already unusual, FLEX exposure should also predict cross-sectional differences in IV, liquidity, and option positioning.

This paper asks four questions:
\begin{enumerate}
    \item Do FLEX events coincide with changes in regular listed-option liquidity, volume, implied volatility, or open interest?
    \item Is the large IV difference between FLEX-exposed and non-FLEX names a single-outlier artifact or a broader selection effect?
    \item Are FLEX events associated with adjusted equity return responses after corporate-action corrections and matched pseudo-events?
    \item Are FLEX signals more useful for directional equity trading or for volatility-carry strategies?
\end{enumerate}

The main empirical design deliberately excludes indices, ETFs, ETNs, and S\&P 500 constituents. The raw FLEX sample includes broad-market and mega-cap instruments whose regular option and equity markets are extremely deep. In those products, daily FLEX consolidations can be informative about institutional risk transfer, but they are unlikely to provide a clean test of whether FLEX activity influences regular option or equity prices. The retained sample is smaller but more appropriate for single-name influence tests.

The paper makes three contributions. First, it constructs a linked FLEX/listed-option/equity event panel from local contract-level and daily files. Second, it shows that the most robust listed-option effect is open-interest formation around FLEX effective dates, not a broad immediate change in IV or spreads. Third, it develops a richer IV analysis showing that FLEX exposure identifies high-IV regimes, especially in the mid-price tier, and that this selection is directly related to the success of volatility-carry strategy prototypes.


\section{Related Work}

\subsection{Option Demand, Inventory, and Implied-Volatility Surfaces}

Classical option pricing begins with complete-market replication \cite{black_scholes_1973, merton_1973}. In that benchmark, demand pressure should not affect option prices except through the underlying and volatility state variables. Empirical option markets depart from this ideal. Bollen and Whaley \cite{bollen_whaley_2004} show that net buying pressure affects implied-volatility functions, and Garleanu, Pedersen, and Poteshman \cite{garleanu_2009} provide a demand-based option-pricing model in which intermediary constraints transmit option demand into prices. Bakshi, Kapadia, and Madan \cite{bakshi_2003} show that individual equity option surfaces contain differential pricing patterns across moneyness and skew, while Carr and Wu \cite{carr_wu_2009} formalize variance risk premia using option-implied variance.

The FLEX setting is related but distinct. We do not observe signed customer demand, and FLEX trades are not the same as daily listed-option order imbalance. Instead, FLEX events reveal customized institutional positioning. The empirical question is whether this revealed positioning is associated with listed-option open interest, volume, or IV regimes.

\subsection{Options, Information, and Equity Predictability}

Easley, O'Hara, and Srinivas \cite{easley_1998} model informed traders choosing between stock and option markets. Pan and Poteshman \cite{pan_poteshman_2006} find that option volume contains information about future stock prices. Ni, Pan, and Poteshman \cite{ni_pan_poteshman_2008} study volatility information trading, and Cremers and Weinbaum \cite{cremers_weinbaum_2010} show that put-call parity deviations predict stock returns. Johnson and So \cite{johnson_so_2012} study the option-to-stock volume ratio as a predictor, while An et al. \cite{an_ang_bali_cakici_2014} connect call and put IV changes to the joint cross-section of stocks and options.

This literature motivates testing directional equity spillovers around FLEX events. It also warns against assuming that every option signal is directional information. Muravyev, Pearson, and Broussard \cite{muravyev_2013} find limited independent price discovery in equity options after accounting for stock-market information. Our evidence is consistent with this caution: FLEX events show a same-day equity CAR difference but no stable multi-day directional signal.

\subsection{Option Returns, Volatility Spreads, and Volatility Carry}

Option return papers are especially relevant for the strategy section. Coval and Shumway \cite{coval_shumway_2001} document expected option-return patterns and losses on zero-beta ATM straddles. Goyal and Saretto \cite{goyal_saretto_2009} show that the cross-section of option returns is related to the difference between historical realized volatility and ATM implied volatility. Bali and Hovakimian \cite{bali_hovakimian_2009} study realized-implied volatility spreads and expected stock returns.

Our event-conditioned short-straddle proxy is a localized version of this broader volatility-spread idea. FLEX exposure identifies names where listed IV is often high; the IV-minus-realized-volatility spread then predicts short-horizon short-straddle P\&L in our sample.

\subsection{Dealer Hedging and Gamma}

Gamma hedging provides a possible mechanism linking options to stocks. Avellaneda and Lipkin \cite{avellaneda_lipkin_2003} study stock pinning near option strikes. Barbon and Buraschi \cite{barbon_buraschi_2021} document intraday momentum and reversal related to dealer gamma imbalances, and Soebhag \cite{soebhag_2023} finds that option gamma predicts stock returns and realized volatility. These papers emphasize that the sign of dealer exposure matters. Our gamma proxy is computed from listed-option gamma and open interest but is not dealer-signed. This limitation helps explain the null gamma-to-next-day-IV result.

\subsection{Event-Study and Difference-in-Differences Design}
\begin{table}[h]
\centering
\small
\setlength{\tabcolsep}{2pt}
\caption{\textbf{Relationship to prior option-market research.} The table maps the main empirical channels in the literature to the corresponding FLEX evidence in this paper.}
\label{tab:literature_map}
\begin{tabular}{p{0.23\textwidth}p{0.24\textwidth}p{0.29\textwidth}p{0.21\textwidth}}
\toprule
\textbf{Literature channel} & \textbf{Representative finding} & \textbf{FLEX analogue in this paper} & \textbf{Evidence} \\
\midrule
Demand pressure and inventory & Net option demand affects IV surfaces \cite{bollen_whaley_2004, garleanu_2009}. & FLEX-exposed names are selected into higher-IV regimes. & \strongcell{Strong for mid-tier IV selection} \\
Option information and equity returns & Option volume and IV signals can predict stock returns \cite{pan_poteshman_2006, johnson_so_2012}. & FLEX flow has weak multi-day equity directionality after matching. & \weakcell{Weak directional evidence} \\
Volatility-spread option returns & IV-RV spreads predict option returns \cite{goyal_saretto_2009}. & IV-RV spread predicts event short-straddle proxy P\&L. & \strongcell{Strong in-sample volatility carry} \\
Dealer gamma hedging & Gamma exposure can affect stock returns and volatility \cite{barbon_buraschi_2021, soebhag_2023}. & Unsigned listed gamma proxy does not predict next-day IV. & \neutralcell{Null without dealer signs} \\
Event-study identification & Matching and clustered DiD improve inference \cite{abadie_imbens_2006, bertrand_2004}. & Matched-control DiD isolates listed-option open-interest effects. & \strongcell{Strong for open interest} \\
\bottomrule
\end{tabular}
\end{table}

The empirical design uses event studies, matching, and fixed-effect DiD. MacKinlay \cite{mackinlay_1997} provides the canonical event-study framework. Abadie and Imbens \cite{abadie_imbens_2006} motivate matching estimators, while Bertrand, Duflo, and Mullainathan \cite{bertrand_2004} emphasize clustered inference in DiD settings. Sun and Abraham \cite{sun_abraham_2021} and Callaway and Sant'Anna \cite{callaway_santanna_2021} caution that dynamic treatment effects in staggered designs can be sensitive to heterogeneous treatment timing. We therefore report paired tests, matched-control DiD, dynamic pre-trend diagnostics, and equity matched-pseudo-event comparisons separately.

\section{Data}

\subsection{Raw Files and Precomputation}

The study window contains 250 trading days from March 31, 2025 to March 27, 2026. Data are sourced from The Options Clearing Corporation (OCC), Cboe Global Markets, and Newmark Risk. The raw files include FLEX consolidations, daily FLEX positions, daily standardized option chains, daily equities, and corporate-action adjustments. The raw standardized option chain is too large for repeated full-memory analysis, so our analysis pipeline uses a precomputed daily underlying panel. For each underlying \(i\) and day \(t\), the precomputation script reads one raw option-chain parquet file and computes:
\begin{align}
\mathrm{Spread}_{it} &= \mathrm{median}_{j \in \mathcal{O}_{it}}\left(\frac{Ask_{ijt}-Bid_{ijt}}{(Ask_{ijt}+Bid_{ijt})/2}\right),\\
\mathrm{IV}_{it} &= \mathrm{median}_{j \in \mathcal{O}_{it}}(IV_{ijt}), \quad 0 < IV_{ijt} < 5,\\
\mathrm{Vol}_{it} &= \sum_{j \in \mathcal{O}_{it}} Volume_{ijt}, \qquad
\mathrm{OI}_{it} = \sum_{j \in \mathcal{O}_{it}} OpenInterest_{ijt},\\
\mathrm{ATMIV}_{it} &= \mathrm{median}_{j: 0.95 \le K_{ijt}/S_{it} \le 1.05}(IV_{ijt}),\\
\mathrm{GEX}_{it} &= \sum_{j \in \mathcal{O}_{it}} \Gamma_{ijt} \times OI_{ijt}.
\end{align}
The script also computes a 25-delta put-skew proxy (both legs restricted to 20--60 days to expiration), term slope, and ATM gamma measures.

\subsection{Filtered Universe}

The raw panel contains 1,541 listed-option underlyings and 75 FLEX underlyings. The analysis universe is
\begin{equation}
\mathcal{U} = \mathcal{U}_{raw} \setminus \left(\mathcal{U}_{index} \cup \mathcal{U}_{ETF/ETN} \cup \mathcal{U}_{\mathrm{S\&P500}}\right).
\end{equation}
The filter removes 477 S\&P 500 constituents, 146 ETFs/ETNs, and 9 indices (632 total exclusions).\footnote{Because exclusion lists are intersected with symbols observed in the listed-option panel or the FLEX consolidation file, one excluded index symbol appears only in the FLEX consolidation file and does not reduce the option panel count ($1,541 - 631 = 910$).} The resulting sample contains 910 listed-option underlyings, 219,915 listed-option aggregate rows, 68 FLEX consolidation rows, 41 FLEX underlyings, and 53 FLEX event dates. The raw files are not modified.

Figures and tables use the price-tier categories high-price tier, mid-price tier, and low-price tier. These are price-tier labels based on median underlying price quantiles (not true market capitalization buckets) and are interpreted as size/price-proxy tiers.

\begin{figure}[htbp]
\centering
\includegraphics[width=\textwidth]{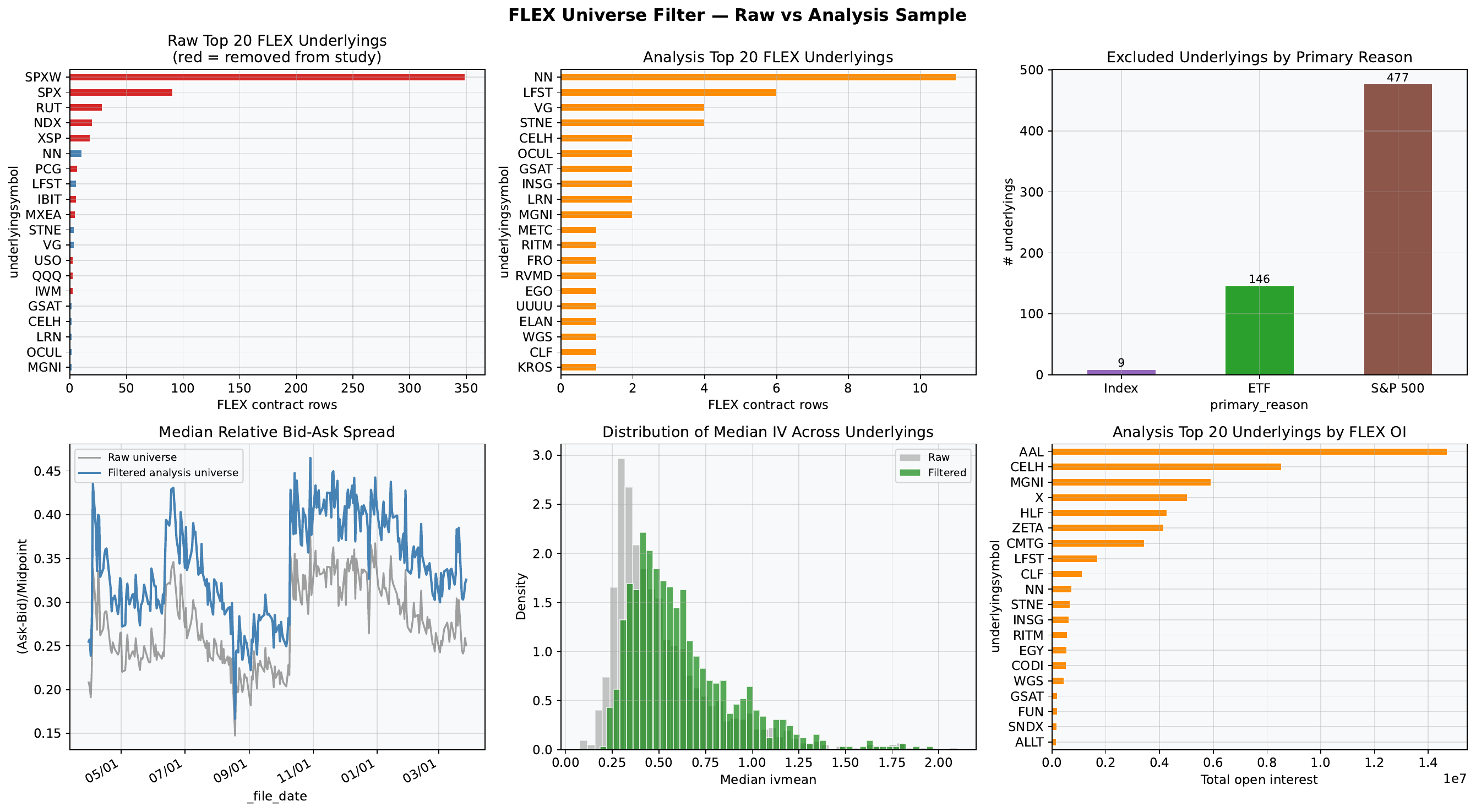}
\caption{\textbf{FLEX sample overview.} The broad raw sample is substantially reduced after removing indices, ETFs/ETNs, and S\&P 500 constituents. This exclusion focuses the empirical design on single-name markets where daily FLEX consolidations are less likely to be overwhelmed by ordinary listed-option and equity-market depth, but it sharply reduces the number of event dates.}
\label{fig:overview}
\end{figure}

\begin{figure}[htbp]
\centering
\includegraphics[width=\textwidth]{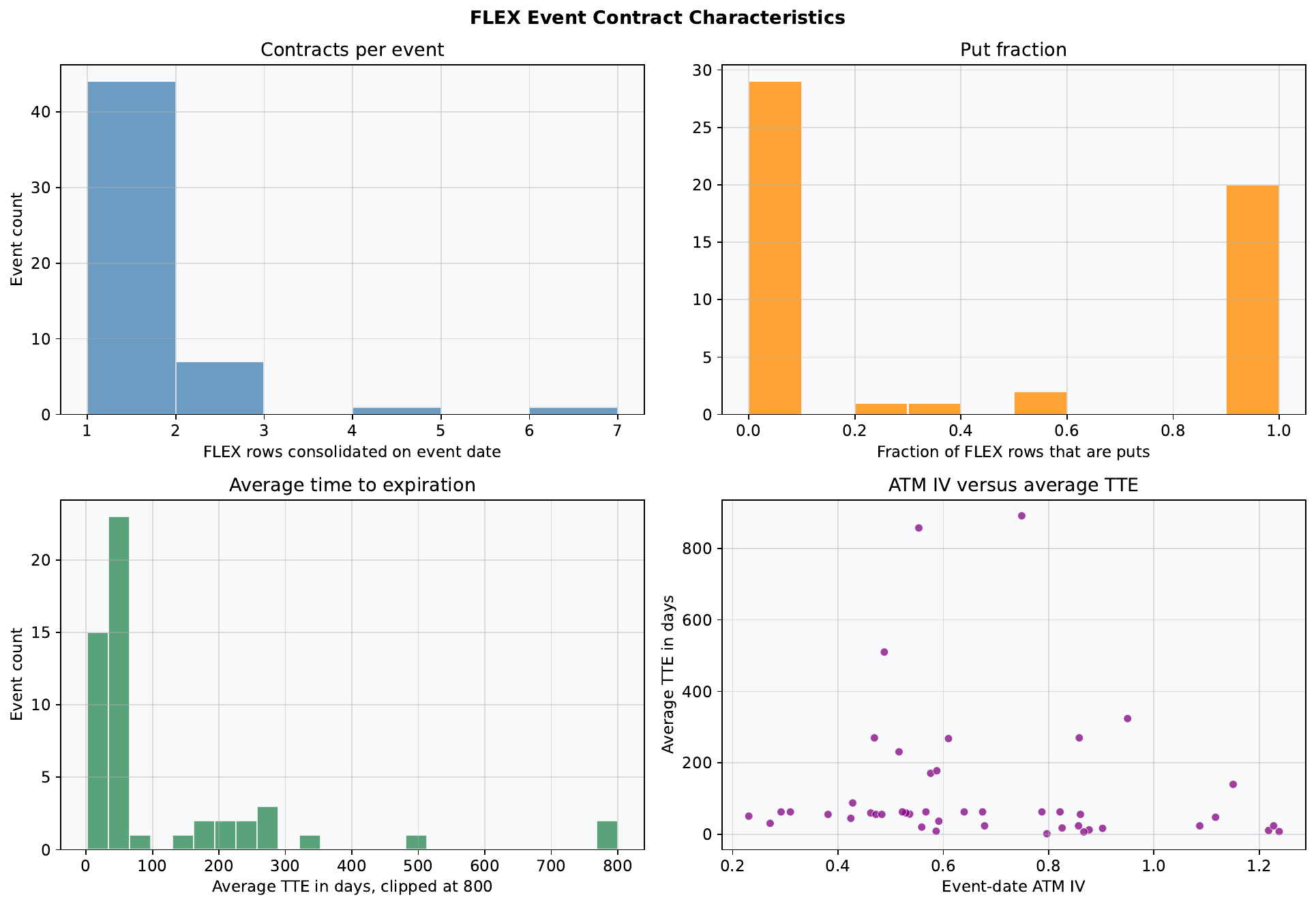}
\caption{\textbf{FLEX contract characteristics in the retained sample.} Most retained events consist of one consolidated FLEX row, but the sample includes multi-row structures such as NN (six contract rows). The put-fraction distribution is bimodal, reflecting predominantly call-only or put-only event days. Average time to expiration is concentrated below three months, with a small number of longer-dated contracts.}
\label{fig:contract_chars}
\end{figure}

\section{Methods}

\subsection{Event Definition}

A FLEX event is an underlying-date pair \(e=(i,t_e)\), where \(t_e\) is the FLEX consolidation effective date. For each event, event time is the trading-day offset
\begin{equation}
\tau(t,t_e) = \mathrm{rank}(t) - \mathrm{rank}(t_e),
\end{equation}
where \(\mathrm{rank}(\cdot)\) indexes the ordered trading calendar. The event window is \(\tau \in [-5,3]\) for listed-option event studies.

\subsection{Paired Event Tests}

For each event \(e\) and listed-option outcome \(y\), define:
\begin{equation}
\Delta y_e = \frac{1}{|\mathcal{T}^{post}_e|}\sum_{\tau=0}^{3} y_{i,t_e+\tau}
- \frac{1}{|\mathcal{T}^{pre}_e|}\sum_{\tau=-5}^{-1} y_{i,t_e+\tau}.
\end{equation}
We report paired t-tests, Wilcoxon signed-rank tests, Cohen's \(d\), and Benjamini--Hochberg multiple-testing corrections across the four listed-option outcomes.

\subsection{Matched-Control Difference-in-Differences}

Controls are selected from non-FLEX underlyings on common support of baseline market conditions: median underlying price, median relative spread, median IV, median log option volume, and median log open interest. The primary DiD model is a staggered-adoption two-way fixed-effects (TWFE) specification:
\begin{equation}
y_{it} = \alpha_i + \lambda_t + \beta\left(Treated_i \times Post_{it}\right) + \varepsilon_{it},
\label{eq:did}
\end{equation}
where \(\alpha_i\) are underlying fixed effects, \(\lambda_t\) are date fixed effects, and $Post_{it} \equiv \mathbf{1}\{t \ge \tau_i\}$ is an indicator set to 1 permanently from the underlying's earliest effective date $\tau_i$ onward. Under this staggered definition, $\beta$ represents the average sustained post-event level shift relative to matched controls across the full post-treatment sample, rather than a narrow windowed effect \cite{callaway_santanna_2021, sun_abraham_2021}. Standard errors are clustered by underlying. Fixed effects are partialled out by iterative two-way residualization.

The dynamic specification replaces the post indicator with event-time dummies:
\begin{equation}
y_{e,\tau} - \bar{y}^{control}_{\tau} =
\alpha_e + \sum_{\tau=-5,\, \tau \ne -1}^{3} \beta_{\tau}\mathbf{1}\{\tau(t,t_e)=\tau\} + u_{e\tau}.
\label{eq:dynamic}
\end{equation}
The omitted period is \(\tau=-1\). Pre-trend diagnostics jointly test \(\beta_{\tau}=0\) for \(\tau<0\).

\subsection{IV Premium Regressions}

The descriptive IV premium is tested both at the symbol level and in a daily panel. The daily panel specification is:
\begin{equation}
IV^{ATM}_{it} = \alpha_t + \delta_{q(i)} + \theta FLEX_i + \varepsilon_{it},
\label{eq:ivpanel}
\end{equation}
where \(\alpha_t\) are date fixed effects, \(q(i)\) is the price-tier group, and \(FLEX_i\) indicates whether the underlying has any retained FLEX exposure. Tier-specific versions omit \(\delta_{q(i)}\) and estimate \(\theta\) within tier. Standard errors are clustered by underlying.

\subsection{Equity Returns}

Daily equity log returns are adjusted for corporate actions:
\begin{equation}
r^{adj}_{it} = \log(P_{it}/P_{i,t-1}) - \log(A_{it}),
\end{equation}
where \(A_{it}\) is the adjustment factor on the ex-date. Abnormal returns subtract a beta-adjusted filtered-universe market return:
\begin{equation}
AR_{it} = r^{adj}_{it} - \hat{\beta}_{i,t-1} r^{mkt}_{t}.
\end{equation}
The cumulative abnormal return from event date \(t_e\) through horizon \(h\) is
\begin{equation}
CAR_{i,e}(h) = \sum_{k=0}^{h} AR_{i,t_e+k},
\end{equation}
with the convention that \(h=0\) is the event-day return. Treated FLEX events are compared with matched pseudo-events from non-FLEX underlyings.

\subsection{Strategy Prototypes}

The trading rules are research prototypes rather than execution-ready strategies. The strategy tests separate a signal proxy from an execution-stressed proxy. The signal proxy asks whether FLEX events identify a high-IV state. The execution stress asks how much of that apparent edge survives bid-ask spreads, simple per-contract fees, and one trading day of signal latency.

For each event \(e\), the one-day straddle-implied move proxy is
\begin{align}
  \widehat{M}_{e} &= 0.8 \times IV_{e} \times \sqrt{1/252},\\
  \Pi^{short,proxy}_{e} &= 0.85 \times \widehat{M}_{e} - |r_{i,t_e+1}|.
\end{align}
The IV-RV spread is \(IV_e - RV^{20}_e\), where \(RV^{20}_e\) is lagged 20-day annualized realized volatility from adjusted equity returns.

The execution-stressed short-straddle proxy uses the event-day median listed-option relative spread \(q_e\), scaled by 1.25 because the aggregate panel does not provide exact ATM-leg quotes. Let \(s_e=\min(0.95,1.25q_e)\). Per-contract fees are converted to return units as \(f_e = 4\times0.65/(100S_e)\), corresponding to two option legs opened and closed on one contract multiplier. The event-close and latency-stressed P\&Ls are:
\begin{align}
  \Pi^{short,exec}_{e} &=
  \widehat{M}_{e}(1-s_e/2) - |r_{i,t_e+1}|(1+s_e/2) - f_e,\\
  \Pi^{short,lag}_{e} &=
  \widehat{M}_{e}(1-s_e/2) - |r_{i,t_e+2}|(1+s_e/2) - f_e.
\end{align}
The latency version uses the close-to-close equity move after a one-trading-day delay. This is intentionally conservative: it tests whether the signal survives if the FLEX consolidation cannot be acted on immediately. Underlying directional strategies subtract the larger of a 10 bps round-trip cost and the observed event-day equity quoted spread when bid/ask quotes are available.

\begin{algorithm}[h]
\caption{Filtered FLEX event-study pipeline}
\label{alg:pipeline}
\begin{algorithmic}[1]
\State Load FLEX consolidations, daily FLEX files, listed-option aggregates, equities, and adjustments.
\State Construct the analysis universe by removing indices, ETFs/ETNs, and S\&P 500 constituents.
\State Define FLEX events as retained \((underlying, effective\_date)\) pairs.
\State Build daily listed-option outcomes: spread, IV, log volume, log open interest, ATM IV, skew, term slope, and gamma proxy.
\State For paired tests, compute pre-event and post-event means within \([-5,-1]\) and \([0,3]\).
\State For DiD, select no-FLEX controls on common support of baseline price, spread, IV, volume, and OI; estimate Eq.~\ref{eq:did}.
\State For dynamic validation, estimate Eq.~\ref{eq:dynamic} and jointly test pre-event coefficients.
\State For equity spillovers, adjust returns for corporate actions and compute matched-control CARs.
\State For strategies, evaluate directional rules and IV-RV short-straddle rules under research-proxy, bid-ask/fee, and one-day-latency assumptions.
\end{algorithmic}
\end{algorithm}

\section{Results}

\subsection{Descriptive Selection}

\begin{figure}[htbp]
\centering
\includegraphics[width=\textwidth]{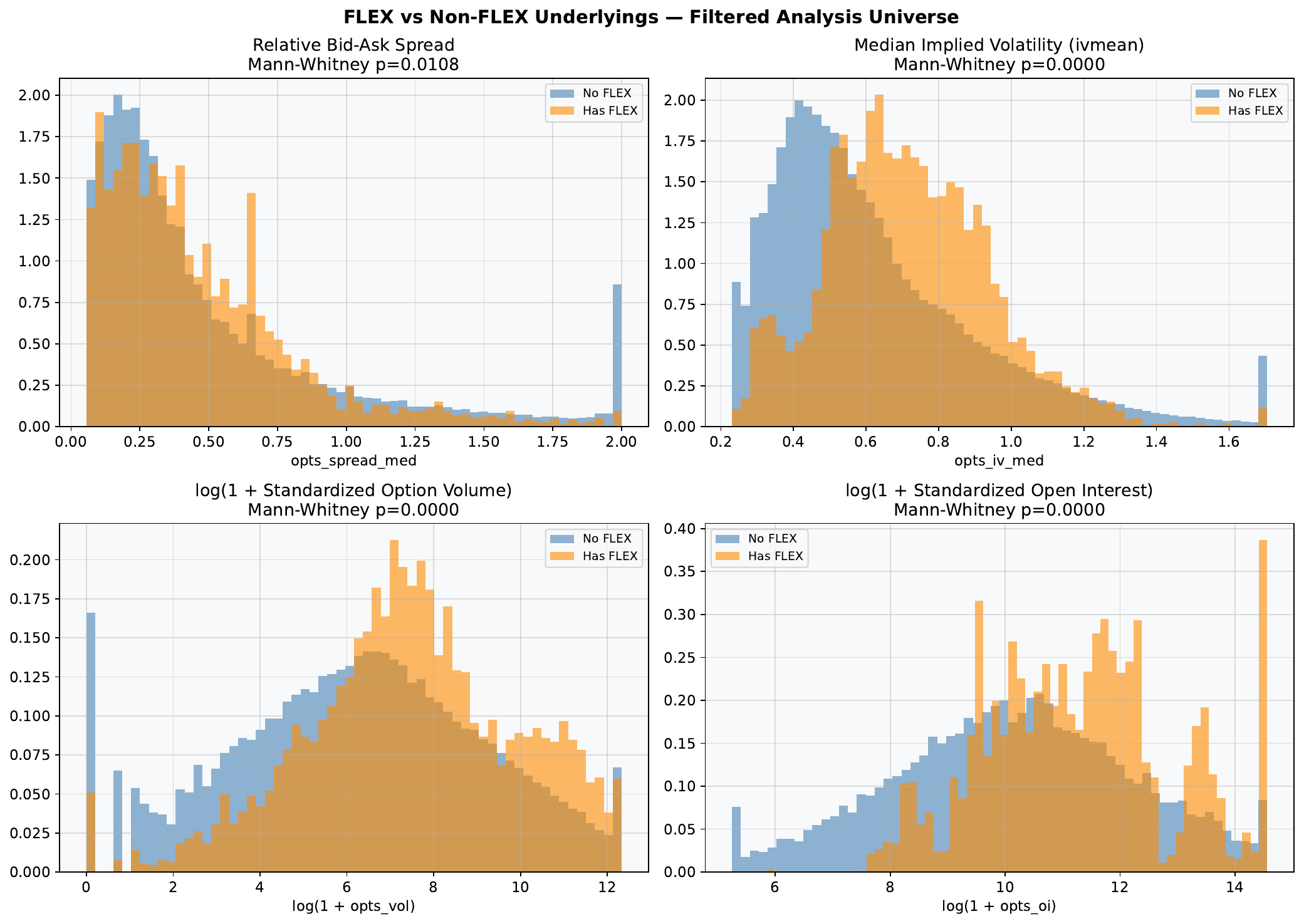}
\caption{\textbf{FLEX exposure and listed-option market conditions.} FLEX-exposed underlyings in the filtered single-name universe have higher listed-option volume, higher listed open interest, and higher listed implied volatility than no-FLEX underlyings. This figure is descriptive and motivates the matched-control and IV-regression analyses; it is not by itself a causal estimate.}
\label{fig:selection}
\end{figure}

Figure~\ref{fig:selection} shows that treatment is highly selected. FLEX-exposed names are active in the listed-option market and have elevated IV. This finding is economically plausible: institutions are more likely to customize exposure in names where standard listed contracts are liquid enough to hedge but incomplete or expensive enough that custom exposure has value.

\subsection{Put-Call Ratios in High-Activity Products}

The exclusion of indices, ETFs, and S\&P 500 constituents is not based on their put-call ratios. It is based on market depth: broad index and ETF option markets are so large that daily FLEX consolidations are unlikely to provide a clean influence test for regular option or equity prices. Nevertheless, these high-activity products provide a valuable descriptive benchmark. We compare actual traded FLEX contract volume and premium spend with same-day listed-option activity for the corresponding products. Crucially, we separate SPX (traditional AM-settled monthly options) from SPXW (PM-settled daily and weekly options) because their contract specifications, expiration cycles, and participant bases differ substantially. FLEX settlement style is identified from the FLEX series prefix, following the OCC consolidation file (e.g., 2SPX maps to AM-settled SPX and 4SPX to PM-settled SPXW).

\begin{table}[htbp]
\centering
\small
\caption{\textbf{FLEX versus listed put-share in high-activity products.} FLEX put share uses actual traded FLEX contract volume from the daily FLEX files, or settlement-price-weighted premium volume. Listed put share uses same-day standardized option volume or midquote-weighted premium volume. PM-settled FLEX products (SPXW, NDXP, RUTW) are benchmarked against the listed SPX, NDX, and RUT chains, respectively. No VIX FLEX trades appear in the sample.}
\label{tab:index_pcr}
\begin{tabular}{lrrrr}
\toprule
Product & FLEX vol. put & Listed vol. put & FLEX prem. put & Listed prem. put \\
\midrule
IBIT & 20.6\% & 35.5\% & 45.3\% & 40.0\% \\
SPX & 51.7\% & 61.6\% & 38.2\% & 35.7\% \\
QQQ & 47.0\% & 53.5\% & 66.3\% & 54.5\% \\
SPXW & 23.0\% & 61.6\% & 14.3\% & 35.7\% \\
IWM & 48.8\% & 61.0\% & 5.1\% & 53.3\% \\
DJX & 49.8\% & 62.3\% & 45.7\% & 40.3\% \\
RUT & 52.9\% & 58.7\% & 52.5\% & 56.6\% \\
NDXP & 6.9\% & 50.7\% & 7.7\% & 44.4\% \\
XSP & 50.8\% & 56.6\% & 87.1\% & 60.6\% \\
RUTW & 6.6\% & 59.2\% & 1.6\% & 58.8\% \\
NDX & 59.2\% & 51.0\% & 48.2\% & 45.7\% \\
MXEF & 10.1\% & 65.5\% & 4.3\% & 73.5\% \\
MXEA & 11.0\% & 60.7\% & 0.5\% & 35.2\% \\
\bottomrule
\end{tabular}
\end{table}

\begin{figure}[htbp]
\centering
\includegraphics[width=\textwidth]{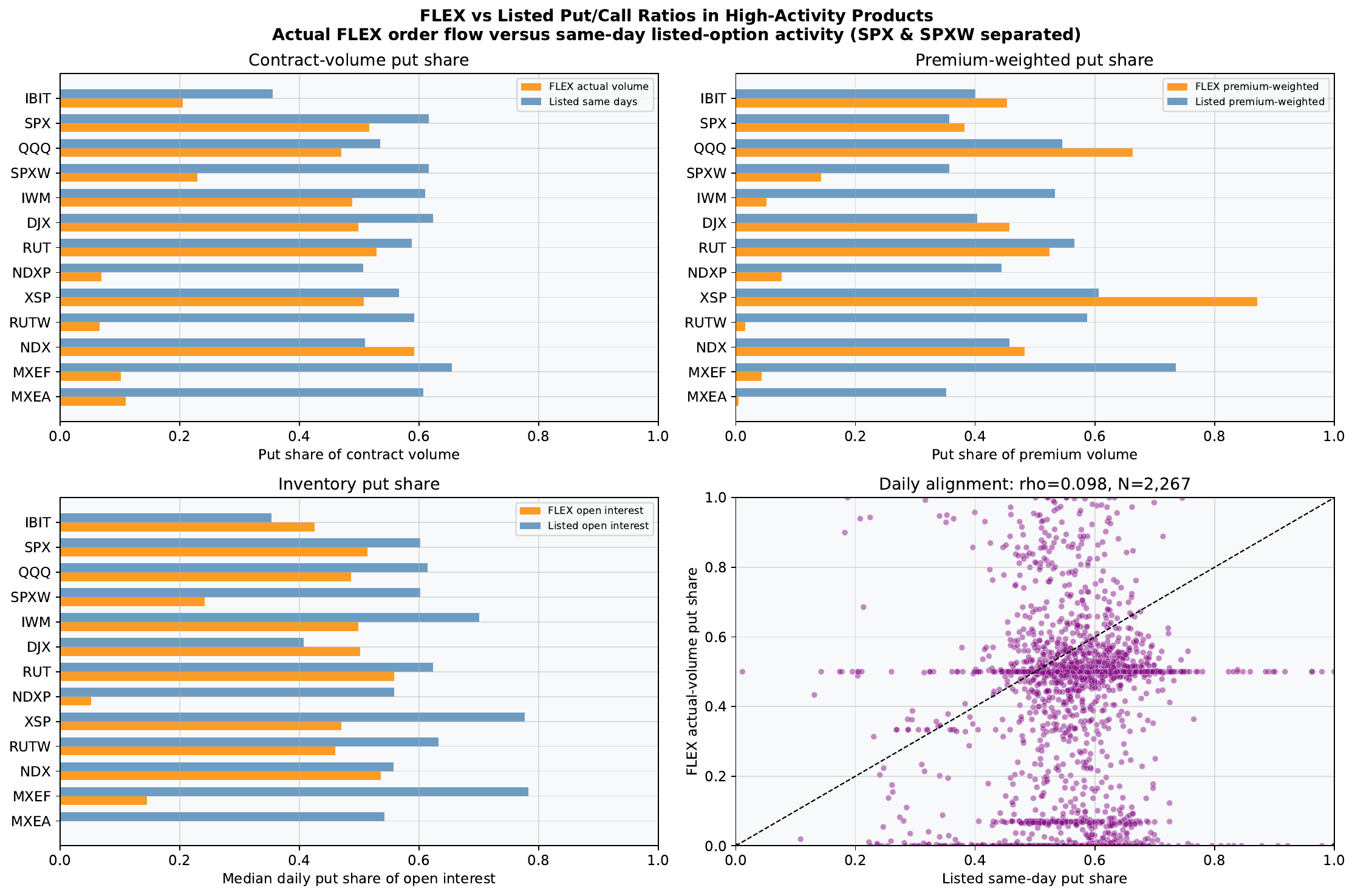}
\caption{\textbf{FLEX versus listed put-call ratios in high-activity products.} Across 2,267 product-days with active FLEX volume, FLEX contract-volume put share is on average 19.7 percentage points below same-day listed put share, and daily co-movement is weak (Pearson \(r=0.063\), Spearman \(\rho=0.098\)). AM-settled SPX FLEX flow has a 51.7\% volume put share versus 61.6\% for listed SPX options, while PM-settled SPXW FLEX flow is strongly call-weighted (23.0\% put share; 14.3\% of premium). NDXP (6.9\%) and RUTW (6.6\%) FLEX flow is similarly call-dominated.}
\label{fig:index_pcr}
\end{figure}

Table~\ref{tab:index_pcr} and Figure~\ref{fig:index_pcr} show that FLEX order flow does not mirror listed-option flow. FLEX volume put share is below the listed benchmark in 12 of the 13 products (NDX is the exception, 59.2\% versus 51.0\%). Separating settlement styles matters: AM-settled SPX FLEX flow is moderately less put-heavy than listed SPX (51.7\% versus 61.6\%), whereas PM-settled SPXW, NDXP, and RUTW FLEX flow is predominantly in calls. Among ETFs, IBIT FLEX flow is call-heavy (20.6\% versus 35.5\% listed). These descriptive benchmarks confirm the active institutional role of FLEX options across index and ETF markets without altering the clean identification strategy of the filtered single-name universe.

\subsection{Listed-Option Event Effects}

\begin{table}[htbp]
\centering
\small
\caption{\textbf{Paired pre/post event tests.} Post is event day through +3 trading days; pre is -5 through -1 trading days. Percent change and Cohen $d$ report directional shifts across 53 event dates. No paired effect survives Benjamini--Hochberg correction at 5\%.}
\label{tab:paired}
\begin{tabular}{lrrrrr}
\toprule
Outcome & Events & Percent change & Cohen \(d\) & t-test p & BH p \\
\midrule
Relative bid-ask spread & 53 & -4.1675 & -0.0615 & 0.3499 & 0.6997 \\
Median IV & 53 & -2.5343 & -0.0424 & 0.0883 & 0.3534 \\
\(\log(1+\mathrm{volume})\) & 53 & 0.2056 & 0.0067 & 0.8641 & 0.8883 \\
\(\log(1+\mathrm{open\ interest})\) & 53 & -0.0389 & -0.0027 & 0.8883 & 0.8883 \\
\bottomrule
\end{tabular}
\end{table}

Paired tests do not provide the main result. The strongest unadjusted paired movement is a small decline in median IV, but it does not survive multiple-testing adjustment. The matched-control DiD is more informative.

\begin{figure}[htbp]
\centering
\includegraphics[width=0.82\textwidth]{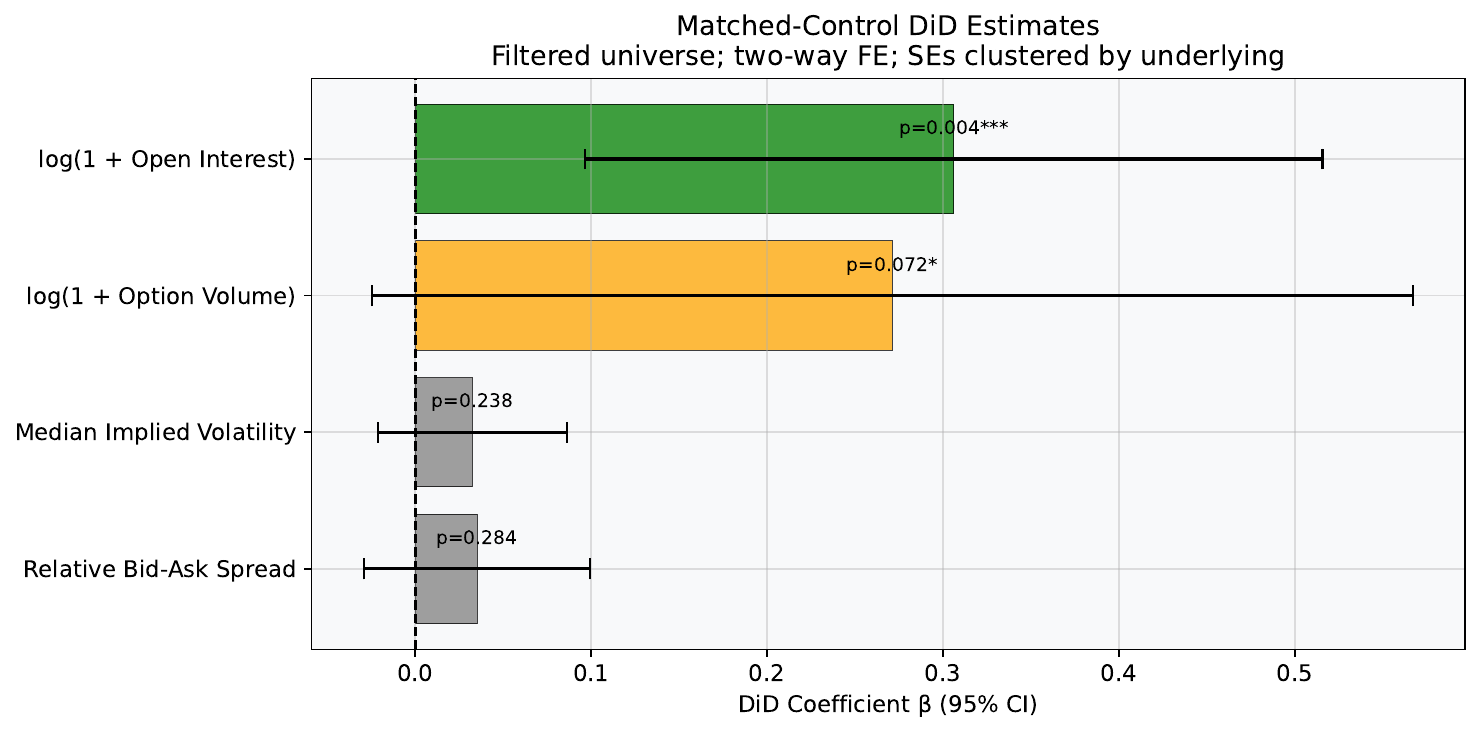}
\caption{\textbf{Matched-control DiD estimates.} Bars show coefficients from Eq.~\ref{eq:did}; error bars are 95\% confidence intervals based on standard errors clustered by underlying. The robust post-event effect is in listed-option open interest.}
\label{fig:did}
\end{figure}

\begin{table}[htbp]
\centering
\small
\caption{\textbf{Matched-control DiD results.} The open-interest coefficient is statistically significant after Benjamini--Hochberg adjustment across outcomes.}
\label{tab:did}
\begin{tabular}{p{0.34\textwidth}rrrrrr}
\toprule
Outcome & Beta & SE & t-stat & p-value & BH p & N \\
\midrule
Relative bid-ask spread & 0.0351 & 0.0328 & 1.0705 & 0.2844 & 0.2844 & 136,211 \\
Median IV & 0.0325 & 0.0275 & 1.1810 & 0.2376 & 0.2844 & 136,142 \\
\(\log(1+\mathrm{volume})\) & 0.2711 & 0.1509 & 1.7964 & 0.0724 & 0.1449 & 136,515 \\
\(\log(1+\mathrm{open\ interest})\) & \strongcell{0.3061} & 0.1069 & 2.8632 & \strongcell{0.0042} & \strongcell{0.0168} & 136,515 \\
\bottomrule
\end{tabular}
\end{table}

The open-interest coefficient of 0.3061 log points is economically meaningful. It indicates that the regular listed-option market accumulates outstanding exposure around FLEX effective dates relative to matched controls. The volume effect is positive but only suggestive after adjustment. This distinction is important: FLEX activity appears more strongly linked to positioning than to immediate regular-option turnover.

\begin{figure}[htbp]
\centering
\includegraphics[width=\textwidth]{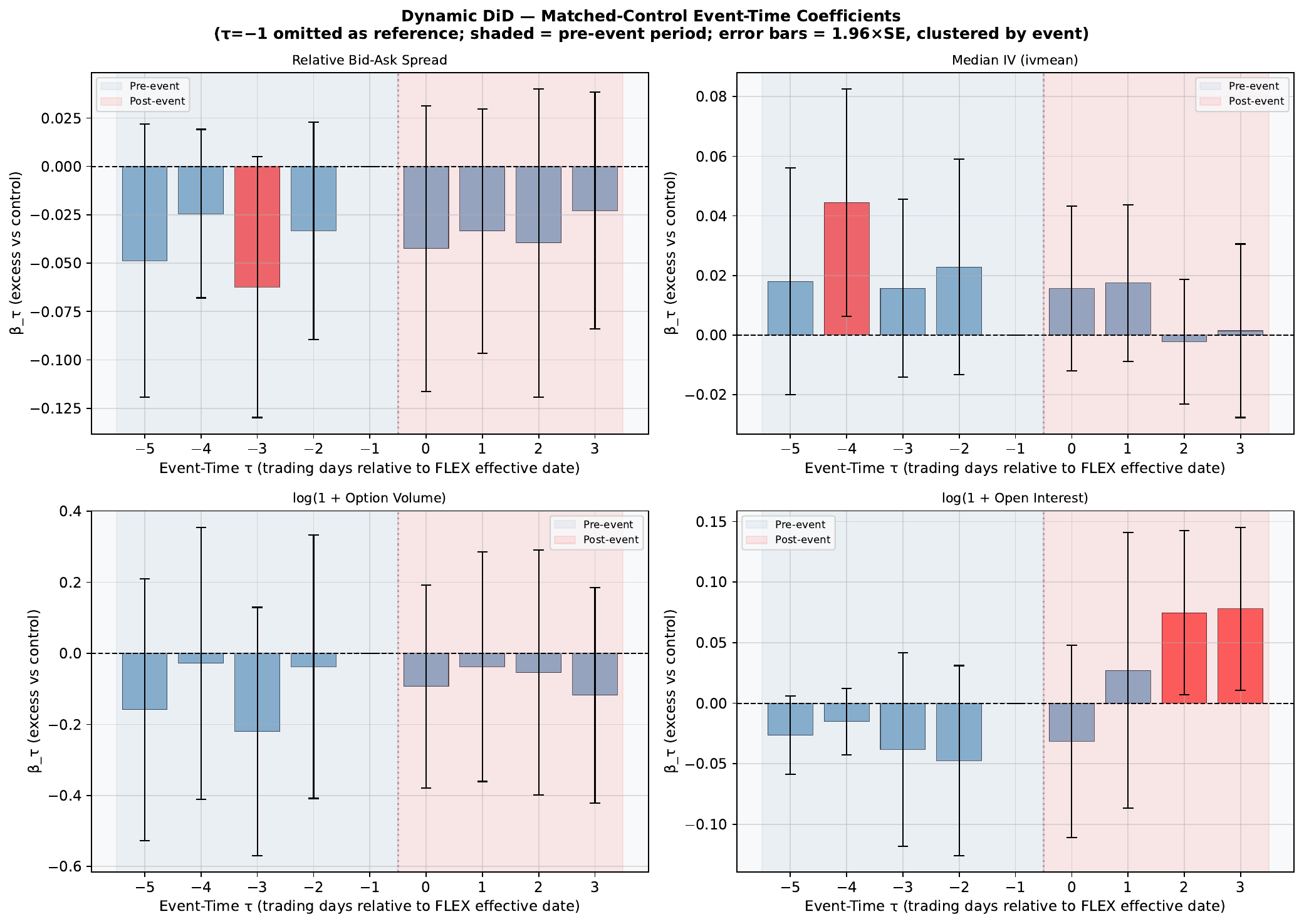}
\caption{\textbf{Dynamic DiD diagnostics.} Event-time estimates use \(\tau=-1\) as the omitted period. Pre-trend tests are acceptable for spread, volume, and open interest, but median IV shows a pre-trend warning. This supports interpreting IV primarily as selection and regime evidence rather than as a clean post-event causal effect.}
\label{fig:dynamic}
\end{figure}

The dynamic pre-trend tests give p-values of 0.1049 for spread, 0.0698 for median IV, 0.6819 for volume, and 0.1997 for open interest. The open-interest result therefore has the cleanest event-time support among the listed-option outcomes.

\subsection{Implied-Volatility Premium}

The IV analysis is one of the central results. FLEX-exposed names have higher listed IV, but this difference must be decomposed into outlier effects, tier composition, and date effects.

\begin{figure}[htbp]
\centering
\includegraphics[width=\textwidth]{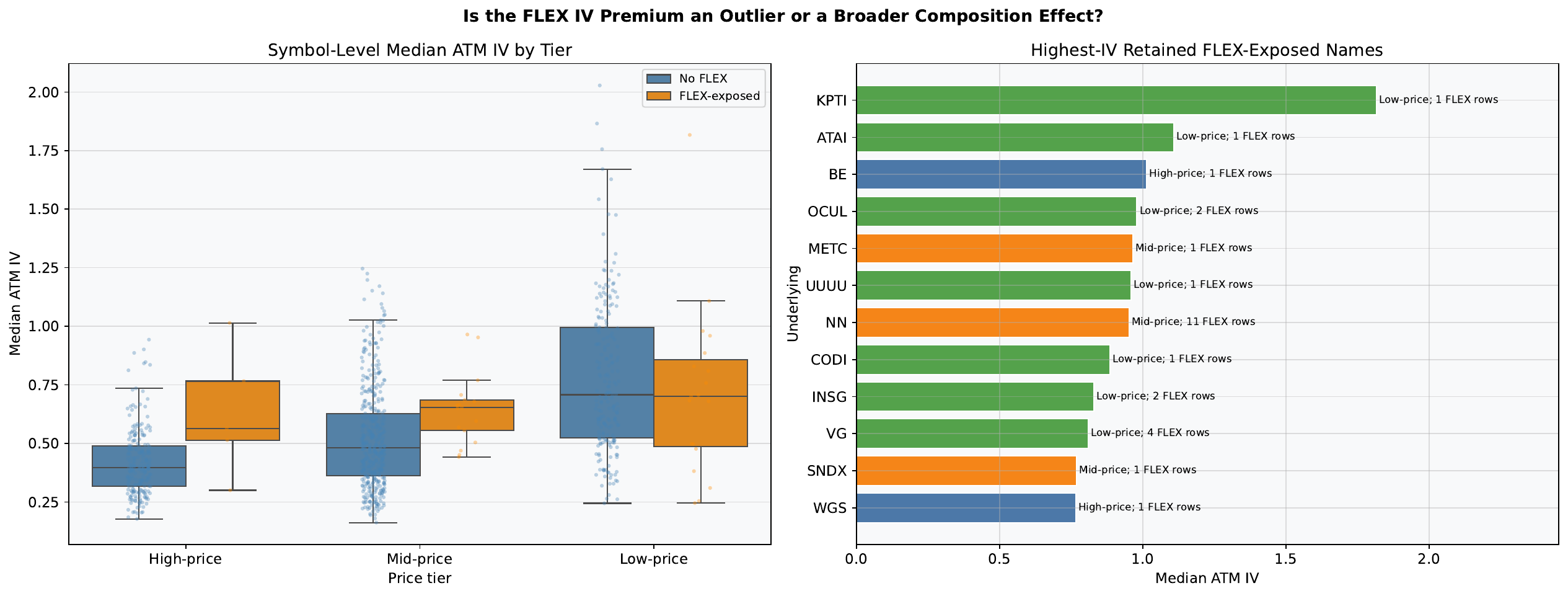}
\caption{\textbf{Symbol-level IV premium diagnostics.} The left panel compares symbol-level median ATM IV by price-tier and FLEX exposure; the right panel lists the highest-IV retained FLEX-exposed names. The mid-price tier result is not driven by a single outlier: the ATM IV gap remains after dropping the highest-IV mid-tier FLEX name.}
\label{fig:ivdiag}
\end{figure}

\begin{table}[htbp]
\centering
\small
\caption{\textbf{Symbol-level IV robustness.} FLEX-minus-control median gaps are computed across symbols with at least 20 observations. ``MW p'' denotes Mann--Whitney U test $p$-value. ``After drop'' removes the highest-IV FLEX-exposed symbol within the tier.}
\label{tab:iv_symbol}
\begin{tabular}{llrrrrr}
\toprule
Price Tier & Metric & FLEX symbols & Control symbols & Median gap & MW p & MW p after drop \\
\midrule
High-price tier & Median listed IV & 5 & 223 & 0.1514 & 0.0603 & 0.2119 \\
High-price tier & ATM IV & 5 & 223 & 0.1659 & 0.0384 & 0.1434 \\
Mid-price tier & Median listed IV & 17 & 437 & \strongcell{0.1366} & \strongcell{0.0053} & \strongcell{0.0122} \\
Mid-price tier & ATM IV & 17 & 437 & \strongcell{0.1723} & \strongcell{0.0025} & \strongcell{0.0063} \\
Low-price tier & Median listed IV & 19 & 209 & 0.0036 & 0.2510 & 0.1175 \\
Low-price tier & ATM IV & 19 & 206 & -0.0074 & 0.3448 & 0.1752 \\
\bottomrule
\end{tabular}
\end{table}

The mid-price tier ATM IV gap is 0.1723 with p-value 0.0025. After removing METC, the highest-IV FLEX-exposed mid-tier name, the gap remains 0.1708 with p-value 0.0063. The high-price tier gap is more fragile because only five FLEX-exposed symbols remain, and the low-price tier gap is not positive because the no-FLEX small-name universe contains many high-volatility speculative names.

\begin{figure}[htbp]
\centering
\includegraphics[width=0.82\textwidth]{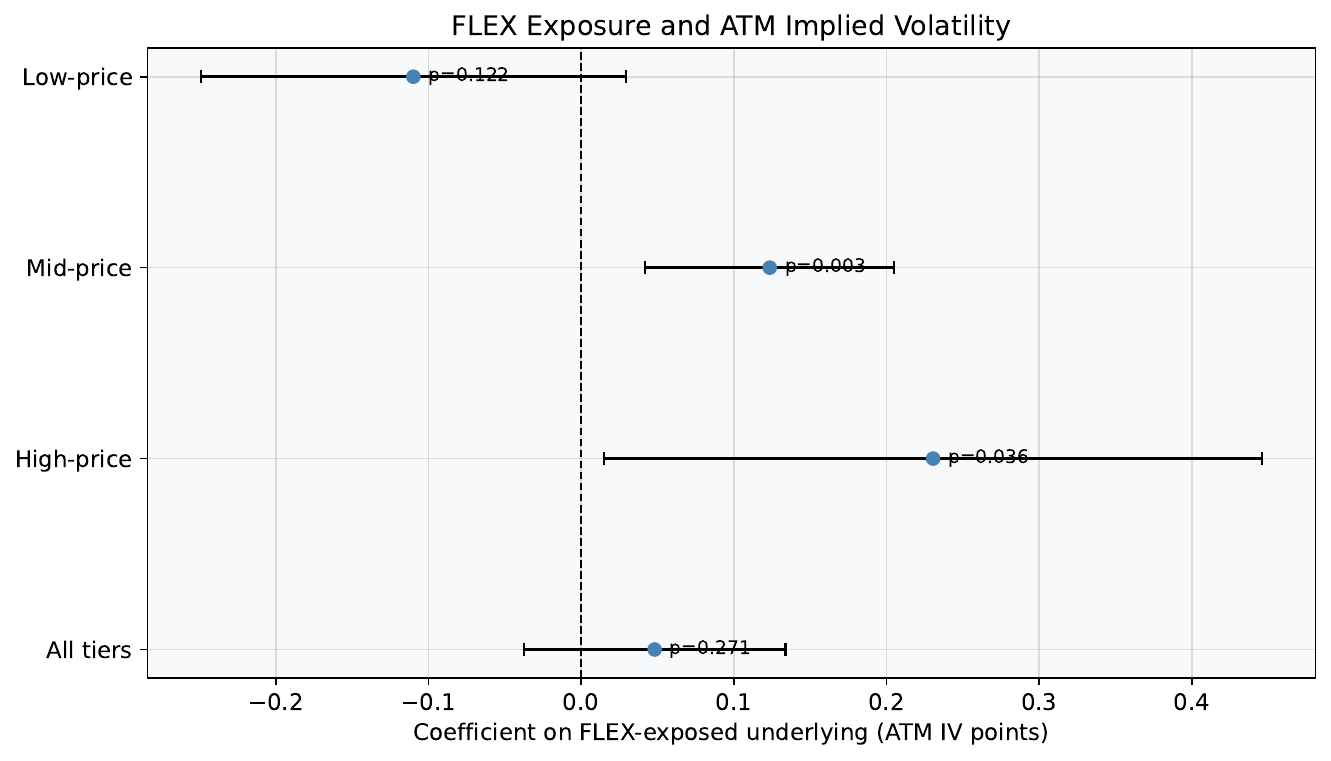}
\caption{\textbf{Date-fixed-effect ATM IV premium regressions.} Points are coefficients on FLEX exposure from Eq.~\ref{eq:ivpanel}; error bars are 95\% confidence intervals using underlying-clustered standard errors. The mid-price tier premium remains large and statistically significant after date controls.}
\label{fig:iv_reg}
\end{figure}

\begin{table}[htbp]
\centering
\small
\caption{\textbf{IV premium panel regressions.} The all-tier model includes date and tier fixed effects. Tier-specific models include date fixed effects. Standard errors are clustered by underlying.}
\label{tab:iv_reg}
\begin{tabular}{p{0.25\textwidth}p{0.25\textwidth}rrrr}
\toprule
Panel & Outcome & FLEX beta & SE & p-value & N \\
\midrule
All price tiers & ATM IV & 0.0481 & 0.0437 & 0.2708 & 188,397 \\
High-price tier & ATM IV & \suggestcell{0.2305} & 0.1100 & \suggestcell{0.0362} & 55,433 \\
Mid-price tier & ATM IV & \strongcell{0.1235} & 0.0415 & \strongcell{0.0030} & 100,874 \\
Low-price tier & ATM IV & -0.1099 & 0.0711 & 0.1219 & 32,090 \\
All price tiers & Median listed IV & 0.0368 & 0.0498 & 0.4603 & 219,287 \\
High-price tier & Median listed IV & 0.1856 & 0.1209 & 0.1247 & 55,562 \\
Mid-price tier & Median listed IV & \strongcell{0.1110} & 0.0398 & \strongcell{0.0052} & 109,892 \\
Low-price tier & Median listed IV & -0.0720 & 0.0911 & 0.4294 & 53,833 \\
\bottomrule
\end{tabular}
\end{table}

The panel regressions clarify the story. Across all tiers, the IV premium is diluted by tier heterogeneity. Within the mid-price tier, however, FLEX exposure is associated with higher ATM IV and higher median listed IV even after date fixed effects. This is the cleanest evidence that FLEX activity identifies a persistent high-IV single-name regime.

\subsection{Event Characteristics and Heterogeneity}

Most retained FLEX events contain a single consolidated contract row. The mean number of FLEX rows per event is 1.283, the median is 1, and the maximum is 6. The mean put fraction is 0.407, reflecting a mixture of call-only and put-only event days. Average time to expiration is 121 calendar days,\footnote{Event sample size varies slightly across specifications depending on data availability: main event panel analyses include $N=53$ events, cost-sensitivity strategy tests drop 1 event ($N=52$) with missing equity return data, and the IV--RV regression drops 1 additional event ($N=51$) with missing 20-day realized volatility data.} but the distribution is right-skewed with several longer-dated contracts.

Cross-event heterogeneity regressions show one strong relationship: log FLEX contract count predicts the event change in \(\log(1+\mathrm{option\ volume})\), with beta 0.1479 and p-value 0.0036. This result complements the DiD evidence by showing that broader or more complex FLEX consolidations are associated with larger regular-option volume changes.

\subsection{Gamma Proxy}

The listed gamma proxy does not predict next-day IV. The coefficient on dealer/listed gamma proxy is \(-7.14\times10^{-10}\) with p-value 0.8063. This should not be read as evidence against dealer hedging more generally. The proxy is not signed by dealer inventory, while gamma-hedging theory depends critically on whether dealers are long or short gamma \cite{barbon_buraschi_2021, soebhag_2023}. In this dataset, gamma concentration is observable but hedge direction is not.

\subsection{Equity Spillovers}

\begin{figure}[htbp]
\centering
\includegraphics[width=\textwidth]{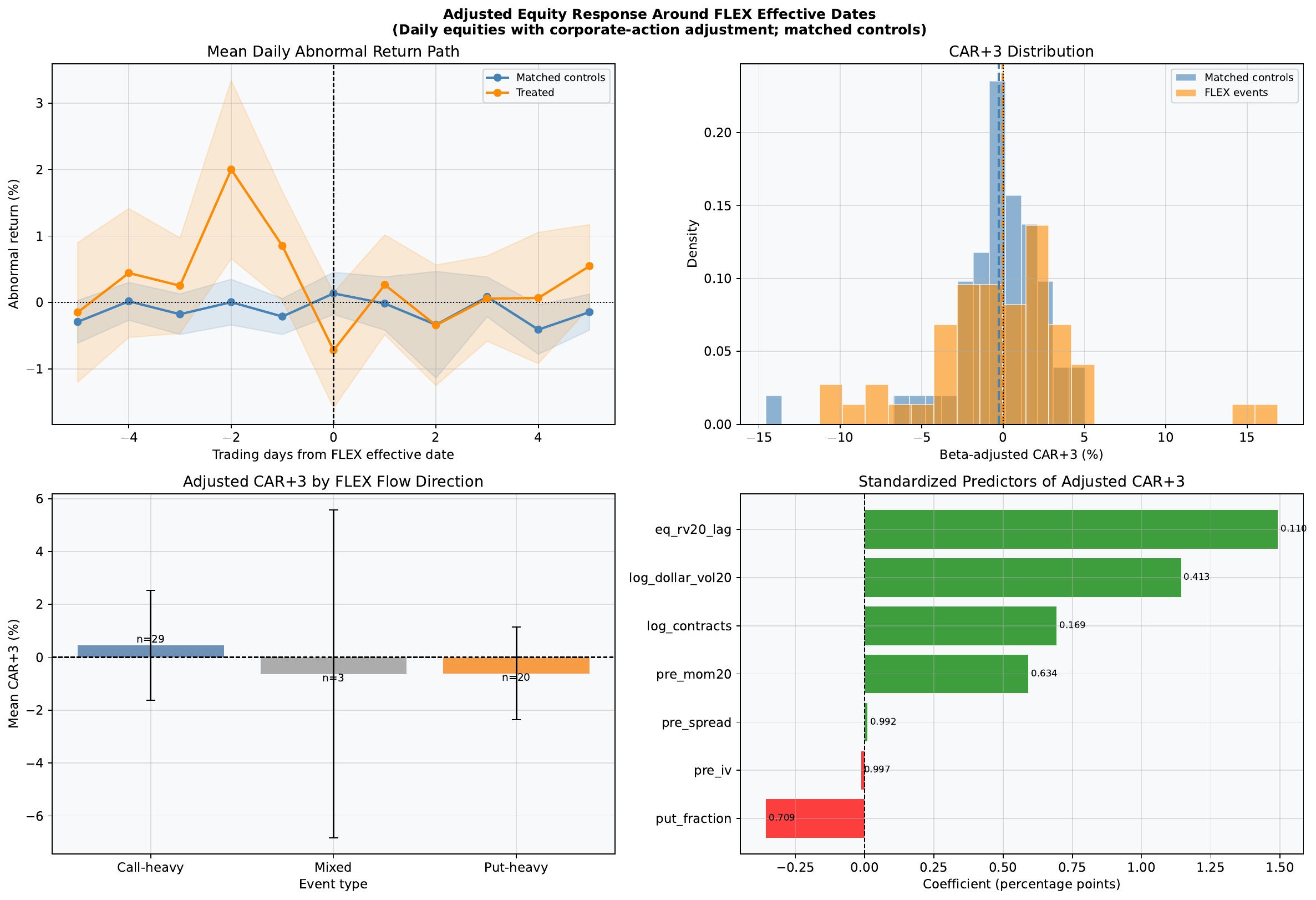}
\caption{\textbf{Adjusted equity response around FLEX events.} Returns are adjusted for corporate actions and beta-adjusted against the filtered equity universe. FLEX events show a same-day matched-control difference, but the effect does not persist at longer horizons.}
\label{fig:equity}
\end{figure}

\begin{table}[htbp]
\centering
\small
\caption{\textbf{Adjusted equity CARs around FLEX events.} Only the same-day horizon ($\text{CAR}_0$) is statistically visible ($p = 0.0442$); subsequent horizons are not statistically significant.}
\label{tab:equity}
\begin{tabular}{lrrrr}
\toprule
Horizon & FLEX mean & Control mean & Difference & MW p \\
\midrule
CAR 0 & \suggestcell{-0.7179\%} & 0.1368\% & \suggestcell{-0.8547\%} & \suggestcell{0.0442} \\
CAR +1 & 0.2665\% & -0.0156\% & 0.2820\% & 0.5262 \\
CAR +3 & -0.0166\% & -0.2663\% & 0.2496\% & 0.9249 \\
CAR +5 & 0.5870\% & -0.8082\% & 1.3953\% & 0.1773 \\
\bottomrule
\end{tabular}
\end{table}

The equity panel contains 220,184 filtered rows across 907 underlyings. Only 1 of the 53 FLEX event windows has a corporate-action adjustment within \(\pm 5\) trading days, so the adjusted-CAR result is not driven by mechanical corporate-action corrections. The same-day effect is economically visible, but the lack of persistence and the absence of significant event-level CAR predictors argue against a broad directional return-predictability claim.

\subsection{Strategy Prototypes}

\begin{figure}[htbp]
\centering
\includegraphics[width=\textwidth]{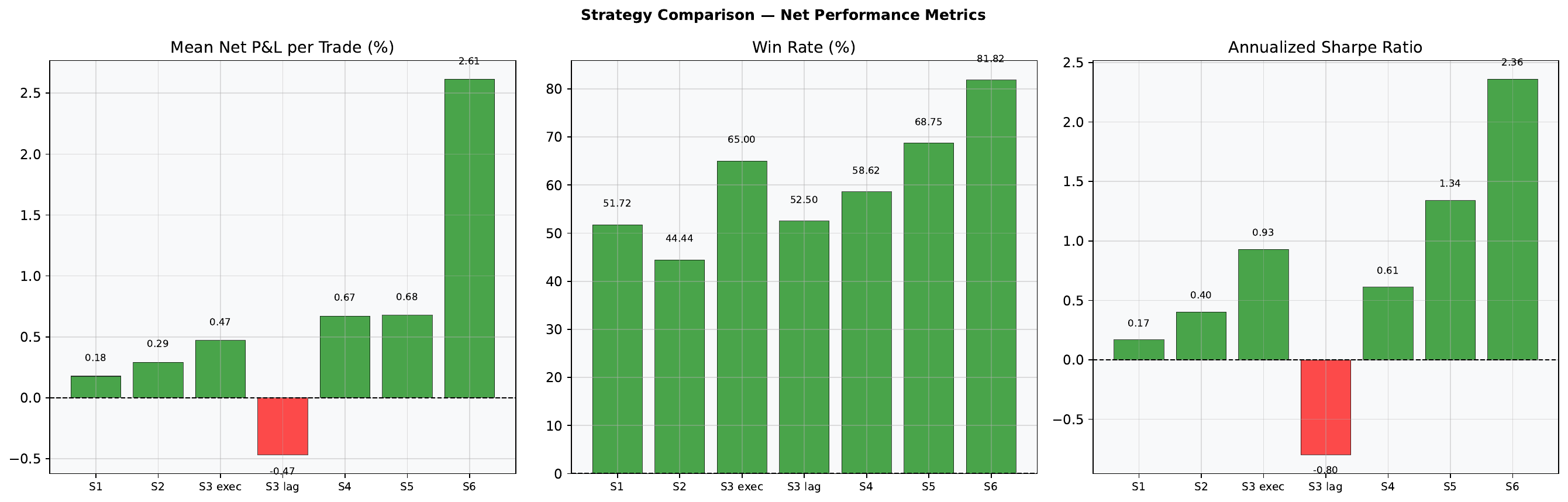}
\caption{\textbf{Strategy comparison.} Directional rules remain statistically weak. The broad rich-IV short-straddle signal is positive after bid-ask and fee stress, but the most attractive in-sample performance comes from the smaller top-IV-RV/liquid subset. The figure should be interpreted as signal screening, not as an execution-ready backtest.}
\label{fig:strategy}
\end{figure}

\begin{table}[htbp]
\centering
\small
\begin{threeparttable}
\caption{\textbf{Strategy prototype performance.} Results are in-sample diagnostics across events. Rows marked ``exec'' include option bid-ask and fee stresses; the latency row enters one trading day after the event signal. Sharpe ratios are annualized by actual trade frequency $\sqrt{N}$ rather than daily trades.}
\label{tab:strategy}
\begin{tabular}{lrrrrrr}
\toprule
Strategy & N & Mean P\&L & Win rate & Sharpe & Max DD & p-value \\
\midrule
Call-pressure long & 29 & 0.1766\% & 51.7\% & 0.17 & -28.97\% & 0.8676 \\
Illiquid put-pressure short & 9 & 0.2914\% & 44.4\% & 0.40 & -3.26\% & 0.6986 \\
Rich-IV short straddle, proxy & 40 & 1.5350\% & 80.0\% & 3.15 & -5.50\% & 0.0031 \\
Rich-IV short straddle, exec & 40 & 0.4704\% & 65.0\% & 0.93 & -8.51\% & 0.3589 \\
Rich-IV short straddle, latency & 40 & -0.4689\% & 52.5\% & -0.80 & -28.72\% & 0.4292 \\
Flow + momentum reversal & 29 & 0.6712\% & 58.6\% & 0.61 & -21.46\% & 0.5464 \\
Liquid rich-IV short straddle, exec & 16 & 0.6767\% & 68.8\% & 1.34 & -4.15\% & 0.2001 \\
Top IV-RV liquid short straddle, exec\tnote{*} & 11 & 2.6144\% & 81.8\% & 2.36 & -3.52\% & 0.0400 \\
\bottomrule
\end{tabular}
\begin{tablenotes}
\small
\item[*] Illustrative double-conditioned subsample (top IV--RV spread and tight listed option spread $\le 75$th percentile) on $N=11$ events, presented for signal screening and requiring out-of-sample confirmation. Sharpe ratio is defined as $(\text{Mean P\&L}/\text{Std P\&L})\times\sqrt{N}$.
\end{tablenotes}
\end{threeparttable}
\end{table}

\begin{figure}[h]
\centering
\includegraphics[width=\textwidth]{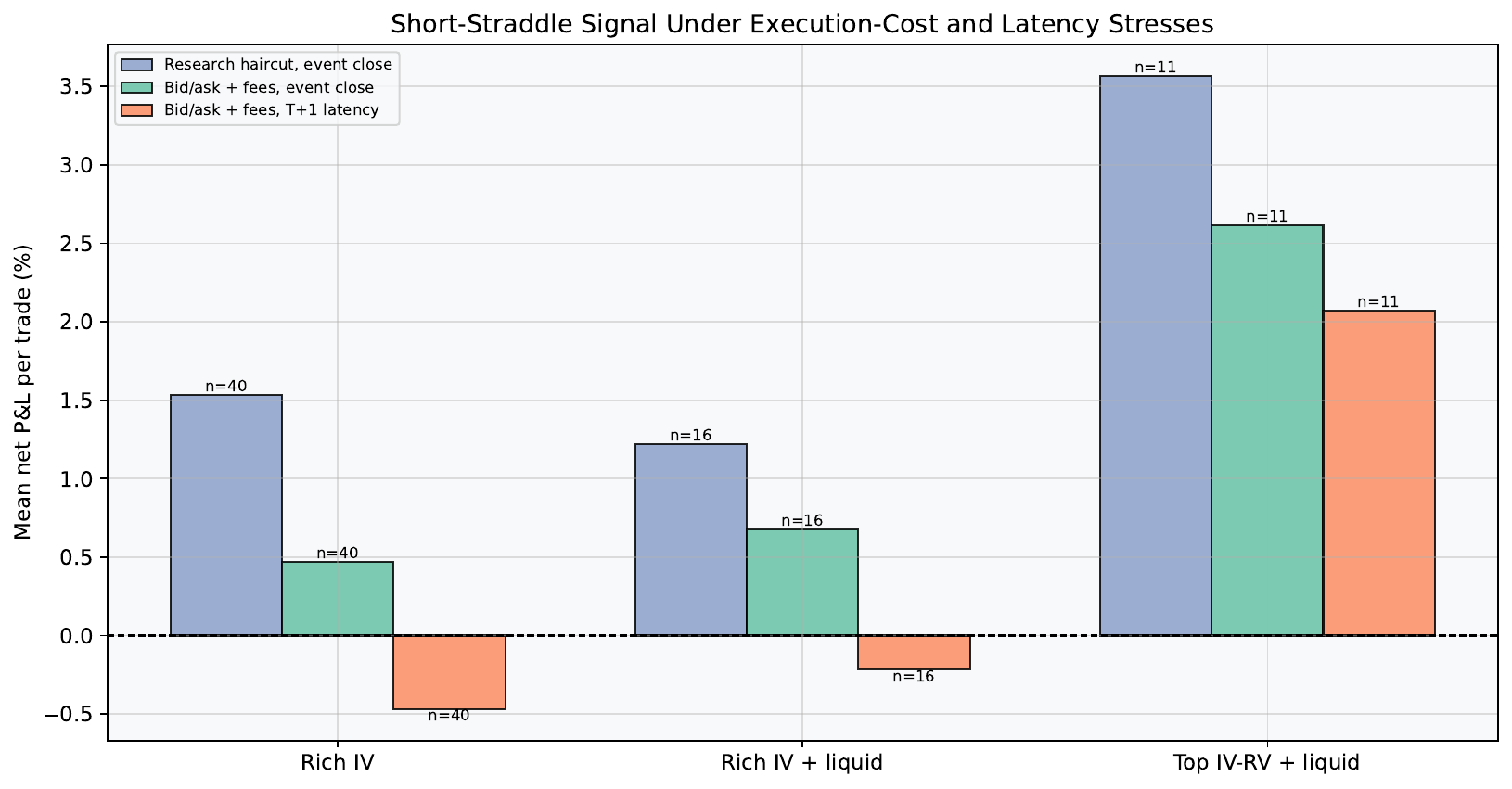}
\caption{\textbf{Short-straddle execution-cost sensitivity.} The broad rich-IV signal loses statistical strength after bid-ask and fee stress and turns negative under a one-day latency stress. The stricter top-IV-RV/liquid rule remains positive under all three cost models, but only over 11 trades.}
\label{fig:cost_sensitivity}
\end{figure}

\begin{table}[h]
\centering
\small
\caption{\textbf{Short-straddle cost and latency sensitivity.} The table compares the original research proxy with bid-ask/fee and latency-stressed variants.}
\label{tab:cost_sensitivity}
\begin{tabular}{p{0.22\textwidth}p{0.28\textwidth}rrrr}
\toprule
Sample & Cost model & N & Mean P\&L & Win rate & p-value \\
\midrule
All events & Proxy, event close & 52 & 1.2858\% & 78.8462\% & 0.0016 \\
All events & Bid-ask + fees & 52 & 0.2978\% & 63.4615\% & 0.4652 \\
All events & Bid-ask + fees + latency & 52 & -0.5397\% & 50.0000\% & 0.2514 \\
Rich IV events & Proxy, event close & 40 & 1.5350\% & 80.0000\% & 0.0031 \\
Rich IV events & Bid-ask + fees & 40 & 0.4704\% & 65.0000\% & 0.3589 \\
Rich IV events & Bid-ask + fees + latency & 40 & -0.4689\% & 52.5000\% & 0.4292 \\
Top IV-RV + liquid & Proxy, event close & 11 & 3.5655\% & 90.9091\% & 0.0164 \\
Top IV-RV + liquid & Bid-ask + fees & 11 & \strongcell{2.6144\%} & \strongcell{81.8182\%} & \suggestcell{0.0400} \\
Top IV-RV + liquid & Bid-ask + fees + latency & 11 & \strongcell{2.0737\%} & \strongcell{81.8182\%} & \suggestcell{0.0306} \\
\bottomrule
\end{tabular}
\end{table}

\begin{figure}[h]
\centering
\includegraphics[width=\textwidth]{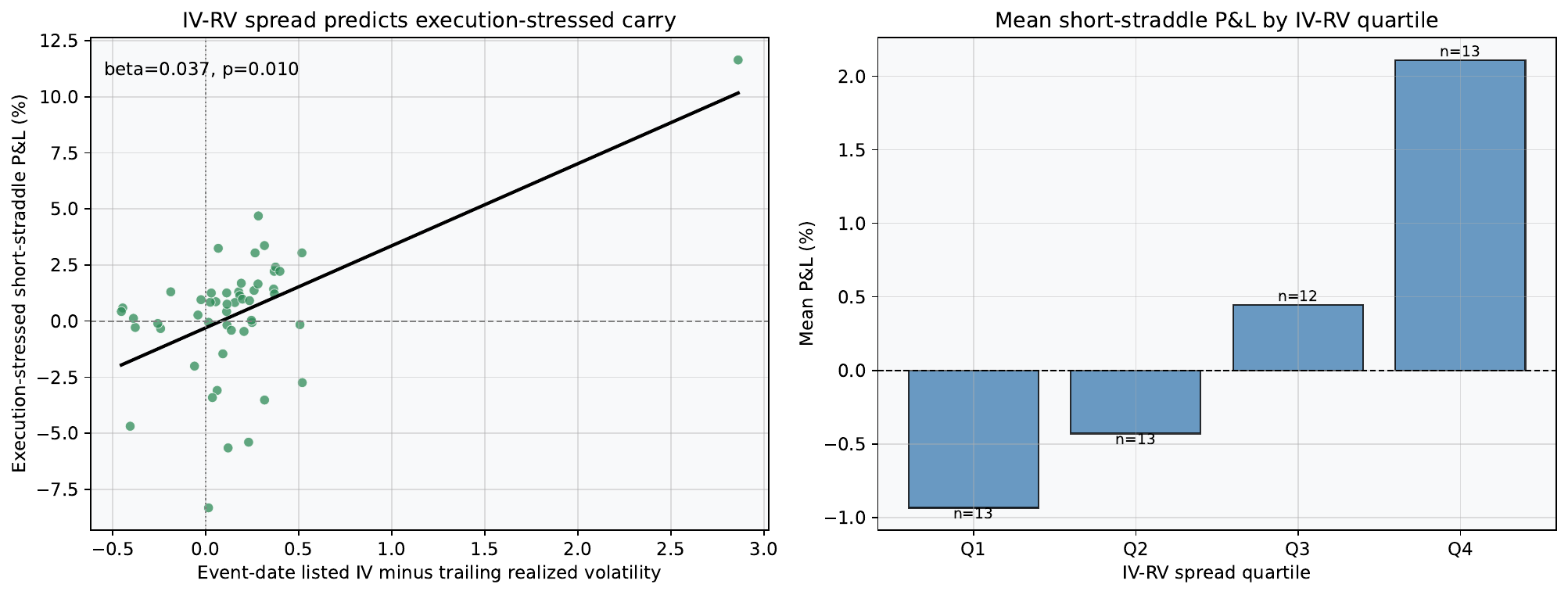}
\caption{\textbf{IV-RV spread and execution-stressed short-straddle P\&L.} The left panel relates event-date listed IV minus trailing realized volatility to bid-ask/fee-stressed one-day short-straddle P\&L. The right panel shows mean P\&L by IV-RV spread quartile. The highest IV-RV quartile has the largest average P\&L even after execution costs.}
\label{fig:ivrv}
\end{figure}

\begin{table}[h]
\centering
\small
\caption{\textbf{Execution-stressed short-straddle P\&L by IV-RV spread quartile.} The monotonic pattern supports the interpretation that the strongest trading signal is volatility carry rather than directional FLEX flow.}
\label{tab:ivrv}
\begin{tabular}{lrrrrr}
\toprule
Quartile & N & Mean IV-RV & Mean P\&L & Win rate & Mean realized move \\
\midrule
Q1 & 13 & -0.2192 & -0.9280\% & 46.15\% & 2.6611\% \\
Q2 & 13 & 0.0834 & -0.4317\% & 53.85\% & 1.9919\% \\
Q3 & 12 & 0.2169 & 0.4530\% & 75.00\% & 1.6796\% \\
Q4 & 13 & \strongcell{0.5756} & \strongcell{2.1068\%} & \strongcell{76.92\%} & 1.9457\% \\
\bottomrule
\end{tabular}
\end{table}

The IV-RV regression is:
\begin{equation}
\Pi^{short,exec}_{e} = -0.0030 + 0.0366(IV_e - RV^{20}_e) + \eta_e,
\end{equation}
with HC3 p-value 0.0100 on the slope. This result directly connects the IV selection evidence to the trading evidence: FLEX events are useful when they identify names where listed IV is high relative to recent realized volatility, but execution costs and latency determine whether that signal is tradable.

\section{Discussion}

The evidence points to a market-structure interpretation. FLEX events help identify a regime in which institutional customization, high listed IV, and regular-option positioning interact. The paper is not arguing that the effective date of a FLEX consolidation mechanically shocks the stock price or mechanically reprices the entire listed-option surface. It argues that FLEX activity is an observable marker of a more latent state: a name has become important enough, unusual enough, or payoff-specific enough that standardized options are no longer the whole story.

The high-activity put-call benchmark adds a complementary point without changing the main identification strategy. In broad products such as the S\&P complex, FLEX flow is abundant enough to study compositionally, but the underlying listed markets are too deep for the same daily event design to isolate price impact. The fact that FLEX put share differs from same-day listed put share shows that FLEX can reveal a different mix of institutional payoff demand than ordinary listed flow. It should be read as an order-flow composition result, not as evidence that index FLEX trades move index option markets.

This framing helps reconcile the results. Open interest responds most clearly because listed options are natural instruments for hedging, offsetting, warehousing, or supplementing a customized FLEX position. IV differences are strong, but they look more like selection into high-volatility regimes than a clean event-date treatment effect. Equity effects are visible only at the event date and fade quickly. The trading evidence is strongest when it is expressed as volatility carry rather than as directional alpha. Taken together, the results describe a layered market: FLEX contracts sit outside the standard chain, but the risk they create or reveal can still be managed through the standard chain and the stock.

The strongest causal-style result is the open-interest DiD. Open interest is a stock variable that accumulates outstanding positions. A robust post-FLEX increase in \(\log(1+\mathrm{OI})\) is consistent with listed options being used to hedge, warehouse, or complement customized FLEX exposures. The volume effect is positive but less robust, which suggests that the listed-market footprint need not appear as a same-day trading burst. This is also why open interest is a more natural outcome than volume in this setting. If a FLEX contract leads dealers or other participants to build offsetting listed positions gradually, the economic footprint may be an inventory stock rather than a one-day flow spike.

The IV evidence is equally important but should be interpreted differently. IV has a pre-trend warning in the dynamic event study, so it is not cleanly identified as a post-event treatment effect. However, the IV premium is real as a selection effect. The mid-price tier remains significant across symbol-level robustness tests and date-fixed-effect regressions. FLEX appears to reveal a subset of single names for which institutions demand customized volatility exposure. This is a useful result even without a causal IV interpretation. In many market applications, selection is the signal: the fact that a name attracts FLEX activity may tell us that the listed surface is expensive, incomplete, or institutionally relevant in a way that is not captured by ordinary volume alone.

The equity evidence is modest. A same-day CAR difference may reflect event-day pressure, hedging, information arrival, or measurement around the consolidation date. The absence of persistence at +1, +3, and +5 days argues against presenting FLEX as a stock-return predictor. This agrees with the cautionary evidence in the broader price-discovery literature \cite{muravyev_2013}. The null persistence result is not a failure of the design; it helps locate the economic channel. In this sample, FLEX is more visible in option-market positioning and volatility-state selection than in sustained stock-market drift.

The strategy evidence resolves the apparent tension, but it also disciplines the claims. Directional rules are weak. The broad rich-IV short-straddle rule is strong as a research proxy, but bid-ask spreads and fees absorb most of the mean, and a one-day latency stress turns it negative. The IV-RV slope remains statistically significant after execution-cost stress, and a narrower top-IV-RV/liquid subset remains positive even with latency, but the subset contains only 11 trades. This is consistent with Goyal and Saretto \cite{goyal_saretto_2009}: the relation between implied and realized volatility can be a tradable option-return signal. In this sample, FLEX events help identify when to look for that spread, while the aggregate data are not sufficient to claim a production-ready strategy.


Several next steps follow naturally. First, future work should collect longer samples and separate equity FLEX, ETF FLEX, and index FLEX as distinct regimes rather than pooling them. Second, exact intraday FLEX auction times and OPRA option quotes would allow cleaner execution analysis and sharper event timing. Third, dealer/customer signs or inventory proxies would turn the gamma test from an unsigned exposure diagnostic into a true hedging-pressure test. Fourth, richer controls for earnings, short interest, borrow costs, institutional ownership, and market capitalization would clarify why certain single names attract FLEX customization. Fifth, a structural model could treat FLEX demand as a latent institutional demand shock that interacts with demand-based option pricing. Finally, the defined-outcome ETF literature suggests a product-level extension: FLEX flows may reveal not only single-name hedging needs, but also the mechanics of packaged payoff products that rebalance or roll in predictable windows.

\begin{table}[h]
\centering
\small
\caption{\textbf{Summary of evidence.} Colors highlight the relative empirical strength of each result in the retained universe.}
\label{tab:summary_evidence}
\begin{tabular}{p{0.26\textwidth}p{0.24\textwidth}p{0.22\textwidth}p{0.18\textwidth}}
\toprule
\textbf{Question} & \textbf{Estimator} & \textbf{Main result} & \textbf{Interpretation} \\
\midrule
Listed-option positioning & Matched-control DiD & \strongcell{\(\beta=0.3061\), BH p=0.0168} & Robust OI spillover \\
IV selection & Date-FE IV regression & \strongcell{Mid ATM IV beta=0.1235, p=0.0030} & Strong selection regime \\
Regular-option volume & Matched-control DiD & \suggestcell{\(\beta=0.2711\), p=0.0724} & Suggestive activity effect \\
High-activity put-call mix & Descriptive FLEX/listed comparison & \neutralcell{Mean FLEX-listed put-share diff=-19.7pp} & Flow composition differs \\
Equity returns & Matched CAR & \suggestcell{CAR 0 diff=-0.85pp, p=0.0442} & Short-lived pressure only \\
Gamma channel & Next-day IV regression & \neutralcell{p=0.8063} & Unsigned proxy is insufficient \\
Trading signal & Exec-stressed IV-RV regression & \strongcell{Slope=0.0366, p=0.0100} & Volatility-carry signal survives costs \\
Trading realism & Cost and latency table & \suggestcell{Broad rich-IV latency mean=-0.47\%; top-IV-RV liquid latency mean=2.07\%} & Signal is promising but sample-limited \\
\bottomrule
\end{tabular}
\end{table}

\section{Conclusion}

FLEX option activity in the filtered single-name universe is best understood as an observable institutional-demand marker. It identifies names with high listed implied volatility, coincides with changes in regular listed-option open interest, and reveals volatility-carry conditions when listed IV is high relative to recent realized volatility. It is not, in this sample, a stable directional equity-return signal or a fully executable trading system.

The central implication is that FLEX should be studied as part of the listed-option market structure, not only as a bespoke product category. FLEX contracts are negotiated outside the regular standardized chain, but their risk is managed in an ecosystem that includes listed options and equities. The paper's evidence is consistent with that ecosystem view. A useful analogy is to think of FLEX as a revealed-preference layer on top of the standardized chain. Standard options show where liquidity is continuously quoted; FLEX shows where some participant wanted a payoff that the standard chain did not naturally provide.


The main limitations are sample size, non-random treatment timing, missing true market-cap data, missing dealer-signed inventory, and aggregate rather than leg-level option execution data. Future work should add market capitalization, institutional ownership, short interest, borrow costs, earnings controls, exact ATM option quotes, intraday quote timestamps, dealer/customer signs, and an out-of-sample design. Even with these limitations, the evidence is enough to show that FLEX is empirically measurable and economically relevant. The paper's role is to establish the first layer of evidence and to define the research object clearly: FLEX is not just an institutional customization tool, but a public signal about where customized nonlinear risk is being created and where that risk may interact with regular markets.

\section*{Data Availability and Source Attribution}
\addcontentsline{toc}{section}{Data Availability and Source Attribution}

Data are sourced from \textbf{The Options Clearing Corporation (OCC)}, \textbf{Cboe Global Markets}, and \textbf{Newmark Risk}. The empirical dataset consists of daily FLEX consolidation records, daily open-interest and position settlement files, standardized listed-option chain aggregates, daily underlying equity prices, and corporate-action adjustments over the study window from March 31, 2025 to March 27, 2026.

\clearpage
\appendix
\section*{Supplementary Information}
\addcontentsline{toc}{section}{Supplementary Information}

\section{Supplementary Method 1: Universe Construction}

The raw data are kept unchanged. The filtered universe is applied inside our analysis pipeline by removing three groups:
\begin{enumerate}
    \item index option products,
    \item ETFs and ETNs,
    \item S\&P 500 constituents.
\end{enumerate}
The exclusion removes 632 underlyings from the listed-option aggregate panel. The purpose is to keep the influence test away from broad-market and mega-cap underlyings whose regular option and equity markets are so deep that daily FLEX consolidations are unlikely to produce an identifiable market-impact signal.

\section{Supplementary Method 2: Precomputed Listed-Option Aggregates}

The raw option files contain approximately 910,000 rows per day on the first study date. The precomputation script processes daily files independently across parallel worker processes and writes one row per underlying per day to \texttt{opts\_agg/}. The aggregation includes median spread, median IV, total volume, total OI, open-interest-weighted gamma exposure, ATM IV, 25-delta skew (put and call legs restricted to a 20--60 day expiration window), term slope, and ATM gamma. This is the empirical unit used by all event-time regressions.

\section{Supplementary Method 3: Strategy Proxy Assumptions}

The strategy tests use simple proxy P\&Ls. Underlying strategies subtract the larger of a 10 bps round-trip cost and the observed event-day equity quoted spread when bid/ask data are available. Liquidity cutoffs for screening (e.g. spread $\le 75$th percentile) are computed on full-sample medians across the study window. Short-straddle strategies are evaluated under three model variants:
\begin{enumerate}
    \item a research proxy applying a flat $0.85$ haircut multiplier to the estimated fair move $\widehat{M}_e$,
    \item an execution-stressed proxy applying proportional bid-ask spread $s_e = \min(0.95, 1.25 q_e)$ haircut factor $(1 - s_e/2)$ and fixed per-contract transaction fees $f_e$,
    \item a one-day latency-stressed proxy entering at $t+1$ close while assuming stale event-day credit $\widehat{M}_e(1 - s_e/2)$.
\end{enumerate}
The research proxy is defined as:
\[
\Pi^{short,proxy}_{e}=0.85 \times \widehat{M}_e - |r_{i,t_e+1}| = 0.85 \times 0.8 \times IV_e \sqrt{1/252} - |r_{i,t_e+1}|.
\]
The execution-stressed proxy uses the median listed-option relative spread because exact ATM-leg quotes are not available in the aggregate panel:
\[
\Pi^{short,exec}_{e}=\widehat{M}_{e}(1-s_e/2)-|r_{i,t_e+1}|(1+s_e/2)-f_e.
\]
Note that the research proxy ($0.85$) and execution-stressed haircut $(1 - s_e/2)$ represent alternative modeling approaches rather than strictly nested haircuts (for tight spreads $s_e < 0.30$, $1 - s_e/2 > 0.85$). The latency variant assumes non-instantaneous quote updates such that entry occurs on day $t+1$ close using stale event-day credit. The implied volatility entering the straddle proxy is the median listed-chain implied volatility from settlement files; no FLEX-derived implied volatilities are computed. Long-straddle candidates use a bid-ask stressed cost and are retained only as a negative baseline. These assumptions are intentionally conservative for signal screening but are not a substitute for intraday option-leg execution modeling.

\clearpage
\section{Supplementary Table 1: Data Coverage}

\begin{table}[H]
\centering
\caption{\textbf{Raw and filtered data coverage.} The filtered analysis universe excludes indices, ETFs/ETNs, and S\&P 500 constituents.}
\begin{tabular}{lrr}
\toprule
Quantity & Raw & Filtered analysis \\
\midrule
Listed-option underlyings & 1,541 & 910 \\
Listed-option aggregate rows & 376,134 & 219,915 \\
FLEX consolidation rows & 628 & 68 \\
FLEX underlyings & 75 & 41 \\
Equity rows & 374,831 & 220,184 \\
Adjustment rows & 8,836 & 3,634 \\
\bottomrule
\end{tabular}
\end{table}

\vspace{1.2em}

\section{Supplementary Table 2: Event Contract Summary}

\begin{table}[H]
\centering
\small
\caption{\textbf{FLEX event contract characteristics.} Statistics are computed over 53 retained FLEX event dates.}
\begin{tabular}{lrrrr}
\toprule
Statistic & Contracts/event & Put fraction & Avg TTE days & Avg moneyness \\
\midrule
Mean & 1.2830 & 0.4072 & 121.3082 & 1.0398 \\
Median & 1.0000 & 0.0000 & 57.0000 & 0.9843 \\
25th percentile & 1.0000 & 0.0000 & 24.0000 & 0.8020 \\
75th percentile & 1.0000 & 1.0000 & 140.0000 & 1.2195 \\
Maximum & 6.0000 & 1.0000 & 891.0000 & 2.1254 \\
\bottomrule
\end{tabular}
\end{table}

\vspace{1.2em}

\section{Supplementary Table 3: Dynamic Pre-Trend Tests}

\begin{table}[H]
\centering
\caption{\textbf{Pre-trend tests for dynamic DiD.} The IV outcome is interpreted mainly as selection evidence because its pre-trend diagnostic is weaker than the open-interest diagnostic.}
\begin{tabular}{lrrl}
\toprule
Outcome & \(\chi^2\) & p-value & Interpretation \\
\midrule
Relative bid-ask spread & 7.66 & 0.1049 & No strong pre-trend warning \\
Median IV & 8.67 & 0.0698 & Pre-trend caution \\
\(\log(1+\mathrm{volume})\) & 2.29 & 0.6819 & No strong pre-trend warning \\
\(\log(1+\mathrm{open\ interest})\) & 5.99 & 0.1997 & No strong pre-trend warning \\
\bottomrule
\end{tabular}
\end{table}

\vspace{1.2em}

\section{Supplementary Table 4: IV-RV Regression}

\begin{table}[H]
\centering
\caption{\textbf{Execution-stressed short-straddle P\&L regression.} The dependent variable is one-day bid-ask/fee-stressed short-straddle proxy P\&L. Standard errors use HC3 robust covariance.}
\begin{tabular}{lrrrr}
\toprule
Term & Coefficient & SE & p-value & N \\
\midrule
Intercept & -0.0030 & 0.0038 & 0.4319 & 51 \\
\(IV-RV\) & 0.0366 & 0.0142 & 0.0100 & 51 \\
\bottomrule
\end{tabular}
\end{table}

\vspace{1.2em}

\section{Supplementary Table 5: Strategy Cost Sensitivity}

\begin{table}[H]
\centering
\small
\caption{\textbf{Short-straddle cost sensitivity.} Broad rich-IV profitability is sensitive to execution assumptions; the top-IV-RV/liquid subset survives the latency stress but has only 11 trades.}
\begin{tabular}{llrrrr}
\toprule
Sample & Cost model & N & Mean P\&L & Win rate & p-value \\
\midrule
All events & Research haircut & 52 & 1.2858\% & 78.8\% & 0.0016 \\
All events & Bid-ask + fees & 52 & 0.2978\% & 63.5\% & 0.4652 \\
All events & Bid-ask + fees + latency & 52 & -0.5397\% & 50.0\% & 0.2514 \\
Rich IV events & Research haircut & 40 & 1.5350\% & 80.0\% & 0.0031 \\
Rich IV events & Bid-ask + fees & 40 & 0.4704\% & 65.0\% & 0.3589 \\
Rich IV events & Bid-ask + fees + latency & 40 & -0.4689\% & 52.5\% & 0.4292 \\
Top IV-RV + liquid & Research haircut & 11 & 3.5655\% & 90.9\% & 0.0164 \\
Top IV-RV + liquid & Bid-ask + fees & 11 & 2.6144\% & 81.8\% & 0.0400 \\
Top IV-RV + liquid & Bid-ask + fees + latency & 11 & 2.0737\% & 81.8\% & 0.0306 \\
\bottomrule
\end{tabular}
\end{table}

\end{document}